\documentclass{aa}
\usepackage[utf8]{inputenc}
\usepackage{graphicx}
\usepackage{txfonts}
\usepackage{verbatim}
\usepackage{siunitx}
\usepackage{hyperref}
\hypersetup{colorlinks, allcolors=blue}
\usepackage{subcaption}
\usepackage{comment}
\usepackage[labelfont=bf]{caption}
\usepackage{tablefootnote}
\usepackage{arydshln}
\usepackage[flushleft]{threeparttable}

\usepackage{color}

\begin{document} 

   \title{Ages, metallicity, and $\alpha$-enhancement of the globular cluster populations in NGC~3311:}
   \subtitle{A stellar population analysis of MUSE co-added 1D spectra}
   \titlerunning{Globular Clusters in NGC~3311}

   \author{Natalie Grasser\inst{1}
    \and Magda Arnaboldi\inst{2}
    \and Carlos Eduardo Barbosa\inst{3}
    \and
    Chiara Spiniello\inst{4}
    \and
    Lodovico Coccato\inst{2}
    \and
    John Trevor Mendel\inst{5,6}
    }

   \institute{University of Vienna, Department of Astrophysics, Türkenschanzstrasse 17, 1180 Vienna, Austria \\
    \email{natalie.grasser@univie.ac.at}
    \and
    European Southern Observatory, Karl-Schwarzschild-Straße 2, 85748, Garching, Germany
    \and
    Universidade de São Paulo, IAG, Departamento de Astronomia, Rua do Matão 1226, São Paulo, SP, Brazil
    \and 
    Sub-Department of Astrophysics, University of Oxford, Denys Wilkinson Building, Keble Road, Oxford OX1 3RH, UK
    \and
    ARC Centre of Excellence for All Sky Astrophysics in 3 Dimensions (ASTRO 3D), Canberra, ACT 2611, Australia
    \and
    Research School of Astronomy and Astrophysics, Australian National University, Canberra, ACT 2611, Australia
    }

\date{}
\abstract
{}
{We aim to investigate the stellar population properties, ages, and metal content of the globular clusters (GCs) in NGC~3311, the central galaxy of the Hydra I cluster, to better constrain its evolution history.}
{We used integral-field spectroscopic data from the Multi-Unit Spectroscopic Explorer (MUSE) to identify 680 sources in the central region of NGC~3311 and extract their 1D spectra. An analysis of these sources in terms of morphologies, radial velocities, and emission lines allowed us to narrow down our selection to $49$ bona fide GC candidates. We split these candidates into two groups depending on their projected distance to the galaxy center ($R$), namely inner ($R\le 20"$) and outer ($R>20''$) GCs. We stacked the extracted 1D spectra of the inner and outer GC populations to increase the signal-to-noise ratios (S/Ns) of the resulting spectra and hence allow full-spectrum fitting. In addition, we also created a stacked spectrum of all GCs in NGC~3311 and one of the two most central GC candidates. Using the code {\tt pPXF}, we performed a stellar population analysis on the four stacked 1D spectra, obtaining mass-weighted integrated ages, metallicities, and $[\alpha/$Fe] abundances.}
{All GCs are old, with ages $\ge13.5$ Gyr, and they have super-solar metallicities. Looking at the color distribution, we find that the inner ones tend to be redder and more metal rich than the outer ones. This is consistent with the two-phase formation scenario. Looking at the full-spectral fitting results, at face value the outer GCs have a larger $[\alpha/$Fe] ratio, which is in line with what is found for the stars that dominate the surface brightness profile at the same radii. However, the values for outer and inner GCs are consistent within the uncertainties. Interestingly, the stacked spectrum of the two most central GCs appears to have the highest metallicity and $[\alpha/$Fe], although with larger uncertainties. They might be associated with the core progenitor of NGC~3311.}
{The careful analysis of the MUSE-extracted 1D spectra for compact sources in the center of NGC~3311 and its halo indicate that a significant fraction (28\%) do display emission lines. Once they are removed, the selected bona fide GC candidates are old ($\ge 13.5$ Gyr) and have super-solar metallicities (slightly larger in the center) and $[\alpha/$Fe] (slightly larger for the outer GCs). Stellar population analyses of the extracted spectra do not support the presence of an intermediate (a few Gigayears) GC population in the central 40 arcsec (10 kpc) radius of NGC~3311.}


\maketitle

\section{Introduction}


Investigating old, massive, early-type galaxies (ETGs) can offer valuable insights into star-formation activity in the early Universe. Considering a two-phase formation scenario for galaxies, which consists of an in situ, high-redshfit star-formation burst and a subsequent time-extended size-growth via accretion (e.g., \citealt{Oser2010, Hilz2012}), massive ETGs are thought to be the result of an extended mass-accretion history. 

In the two-phase formation scenario, the first phase, also called the in-situ phase, encompasses fast and intense processes that form the central bulk mass of the galaxy from large amounts of gas. In the second phase, namely the accretion phase, accretion, in the form of mergers and gas inflows, is responsible for further size growth and structural evolution. The first stage of formation produces a so-called compact progenitor (CP), which usually does not survive until the present day. In fact, mergers (especially major mergers) can cause a mixing of the CP and accreted populations, and they may often trigger a subsequent in situ star-formation episode 
\citep{Pulsoni2021}. However, sometimes CPs can survive almost untouched, either in the central core of a massive galaxy, when it experienced only minor and mini mergers \citep{Barbosa2021}, or even in isolation when they completely miss the second phase of the two-phase formation scenario (relics; see, e.g., \citealt{Trujillo09, FerreMateu17, Spiniello21a, Spiniello21b}). CPs that survive undisturbed until the present offer valuable insights into the star-formation activity at early epochs due to them preserving their initial conditions.

Studying evolved stellar populations, such as those found in globular clusters (GCs) located in the halos of  massive galaxies, is particularly insightful for understanding the conditions during their formation. 
In fact, GCs form out of the in-falling gas of their host galaxy, and hence they can preserve information on the formation conditions of the galaxy, similarly to CPs. If a CP of a galaxy is preserved, then it would be reasonable to assume that the associated GCs are also preserved, since the galaxy had not had mergers that changed the CP significantly, as these were formed after the initial in situ phase. Investigating the color and metallicity distribution of GCs in massive ETGs can therefore provide information on the accretion history of the galaxy and early stellar formation in the Universe. 

In this context, the giant elliptical galaxy NGC~3311, which is the central galaxy as well as the brightest cluster galaxy (BCG) of the Hydra I (Abell 1060) cluster, is an interesting target with an extended GC population (e.g., \citealt{Hempel05, Misgeld2011, Hilker2018}). 
NGC~3311 has a high surface brightness central component surrounded by an extended and diffuse stellar halo \citep{Barbosa2018}; at radii larger than $50'',$  an additional, off-centered envelope (whose centroid is shifted by about $50''$  to the north-east) is present \citep{Arnaboldi2012}. The high surface brightness component identified in \citet{Barbosa2018} is indeed an example of a preserved compact progenitor, which is shown to be very old and metal-rich in the stellar population studies by \citet{Barbosa2021}.
The presence of multiple components in the surface brightness profile of NGC~3311 and of a rising velocity dispersion profile (\citealt{Ventimiglia2011}; \citealt{Hilker2018}) indicates that the galaxy has had an extended accretion history. The accreted stars at large radii onto this central cluster galaxy came from rather massive and quenched satellites, given the rising [$\alpha$/Fe] radial profile (see \citealt{Barbosa2021}). Footprints of this extended mass accretion may be connected with the gas/dust disc or ring in the central region \citep{Richtler20} and to previous claims of the presence of an intermediate-age GC population in the central regions of NGC~3311 \citep{Hempel05}.  

The availability of deep  observations for NGC~3311 enable us to investigate the age and metallicity of GCs associated with this central cluster galaxy. Their age,  metallicity, and $[\alpha$/Fe] distribution may offer insights into the different phases of the galaxy mass assembly.

The paper is organized as follows. 
In Section~\ref{sec:data}, we briefly describe the MUSE data we used. In Section~\ref{sec:methods}, we provide a detailed description of the routines we used to identify GC candidates. In Section~\ref{sec:catalog}, we present the final GC catalog, study the color distribution, and then produce stacked 1D spectra, for the all, inner, and outer GC populations, plus an additional fourth stacked 1D spectrum from the innermost GCs. In Section~\ref{sec:stelpop}, we perform the stellar population analysis on these deep spectra, and, finally, we discuss our findings and present our conclusions in Section~\ref{sec:conclusion}.

\section{Data}
\label{sec:data}
To investigate the GC population in NGC~3311, we used spectral data provided by the Multi-Unit Spectroscopic Explorer (MUSE) spectrograph (\citealt{Henault2004, MUSE2010}), which is mounted at the Nasmyth focus of the UT4 8m telescope of the Very Large Telescope (VLT) in Chile. The MUSE observations were obtained  under the ESO program 094.B-0711A (PI: Arnaboldi).  They were divided into four fields: A, B, C, and D, with the first three positioned along the major axis of NGC~3311 at increasing galactocentric distance, to sample the offset extended envelope \citep{Arnaboldi2012, Barbosa2018}, and Field D targeting HCC007, a spectroscopically confirmed member of the Hydra I cluster (see Fig.~1 in \citealt{Barbosa2018}). 
The data are in the form of datacubes, with the 2D sky position and a spectrum in the  $4650-9300 \AA$ wavelength range in each "spaxel". 
For details on the data reduction and analysis, we refer the reader to \citet{Barbosa2018}, while for a detailed stellar population analysis, which allowed us to prove that NGC~3311 has a preserved CP in its core, we refer the reader to \citet{Barbosa2021}.

\section{Methods}
\label{sec:methods}

In this section, we describe the routines and methods we used to extract the spectra of all  point-like sources from the datacubes, and then to identify, among them, the most reliable GC candidates. In particular, for this purpose, we applied three different selection criteria based on morphological, kinematical, and spectroscopical characteristics of the extracted sources, which we describe in detail in the next subsections.

\subsection{Source identification}

We started by creating a signal-to-noise ratio (S/N) map for each of the fields in order to identify and extract the positions of sources in NGC~3311. We collapsed each single datacube across the wavelength axis to produce a white-light image from the flux and a noise image from the variance. Then, we divided them to create a S/N 2D image.
For this, we only used data in the wavelength range of $4800-7000 \,\AA$ to avoid contamination from telluric lines at redder wavelengths. Figure~\ref{fig:fields_S/N} visualizes the S/N map of the MUSE data for all four fields.

\begin{figure*}[t!]
        \centering
        \begin{subfigure}[b]{0.475\textwidth}
            \centering
            \includegraphics[width=\textwidth]{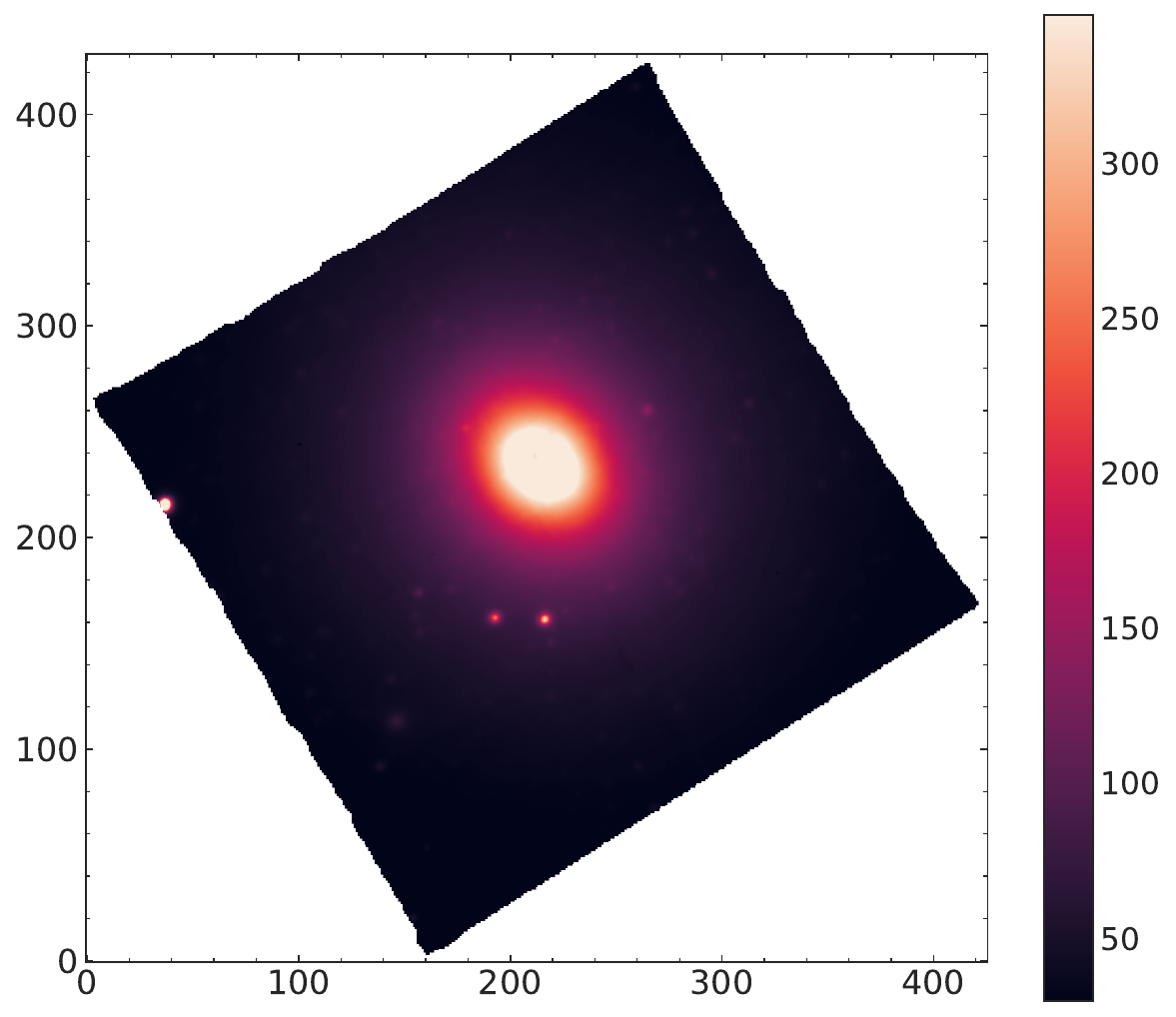}
            \caption{Field A}           \end{subfigure}
        \hfill
        \begin{subfigure}[b]{0.475\textwidth}  
            \centering 
            \includegraphics[width=\textwidth]{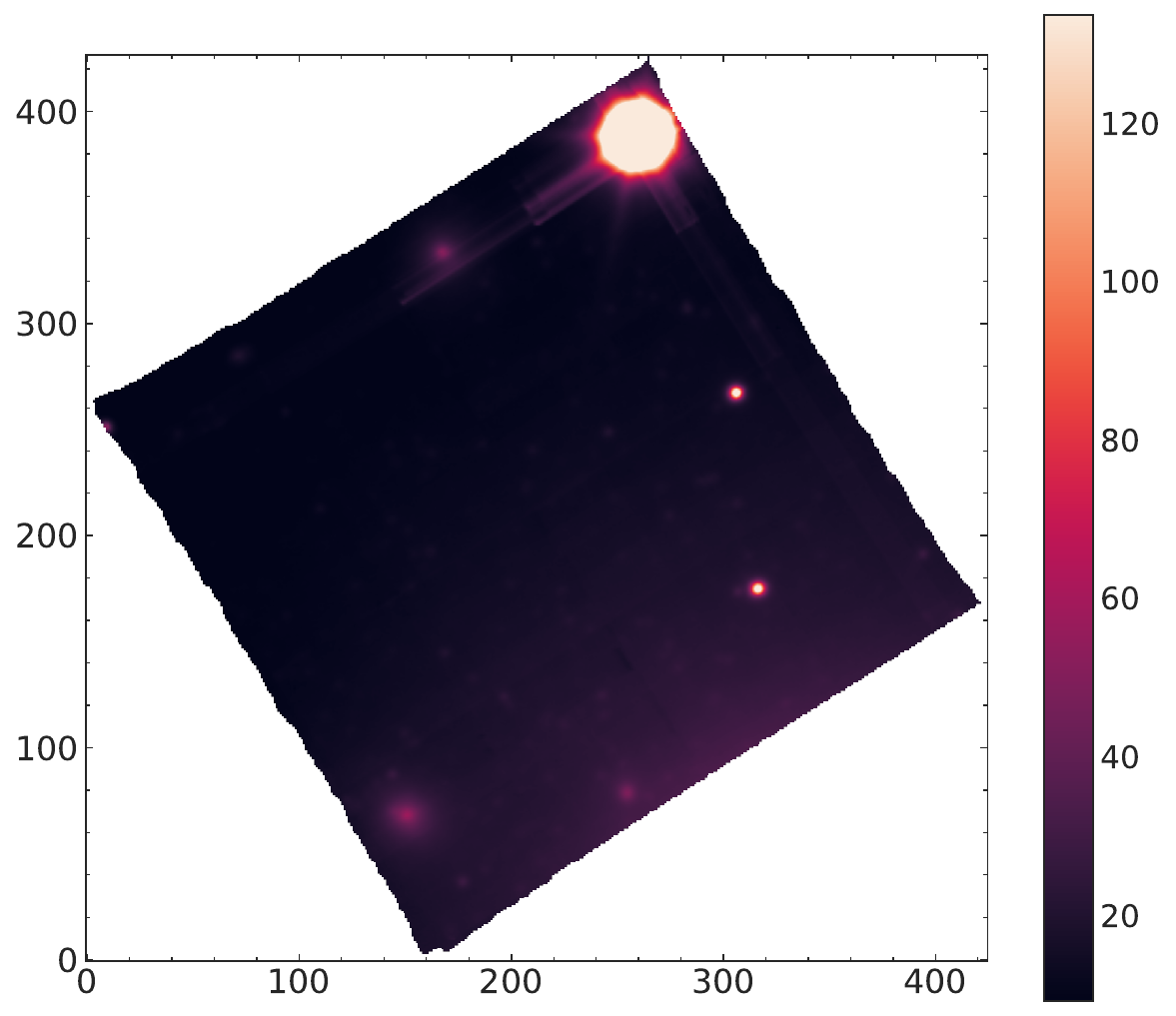}
            \caption{Field B}    
        \end{subfigure}
        \vskip\baselineskip
        \begin{subfigure}[b]{0.475\textwidth}   
            \centering 
            \includegraphics[width=\textwidth]{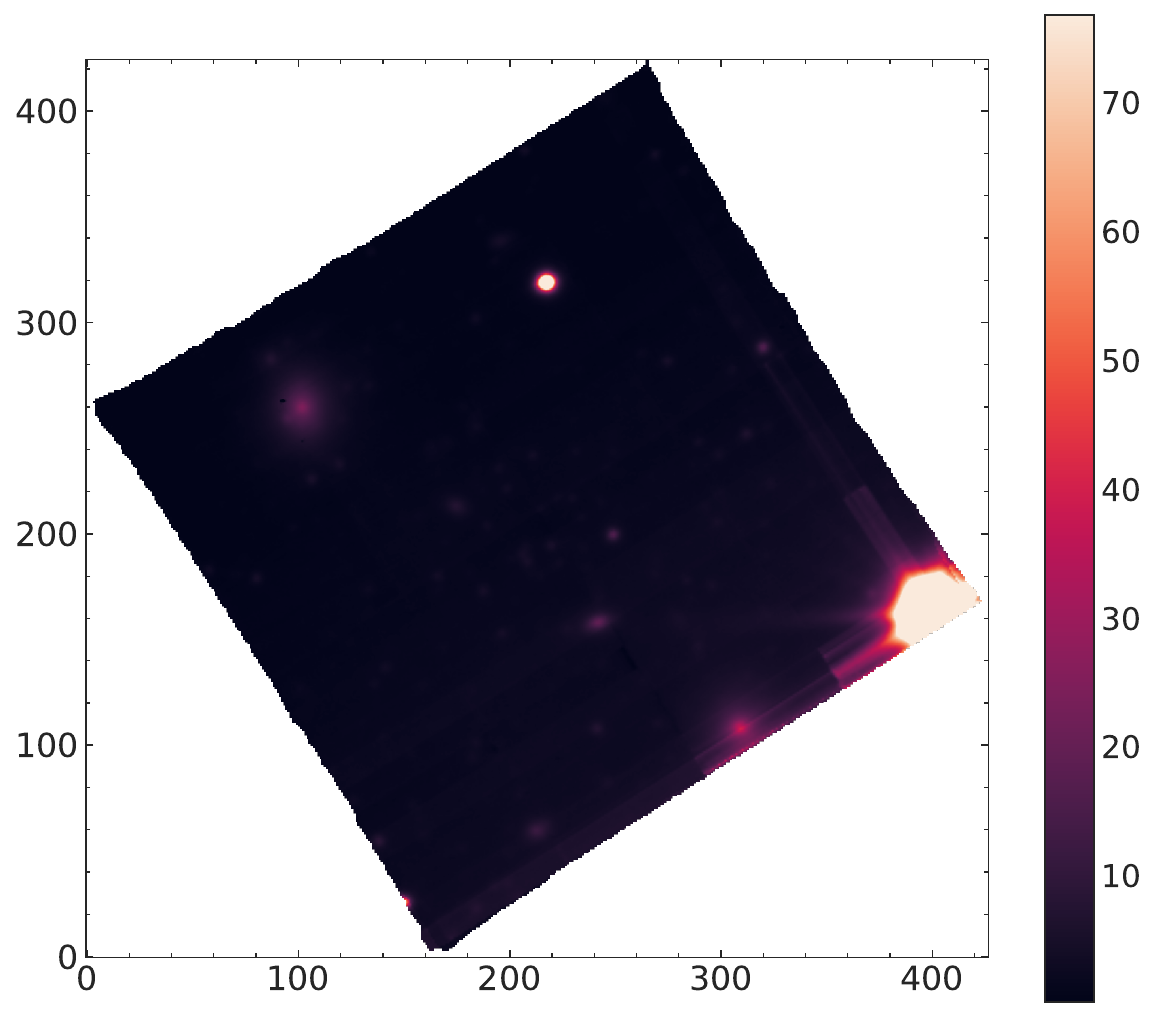}
            \caption{Field C}   
        \end{subfigure}
        \hfill
        \begin{subfigure}[b]{0.475\textwidth}   
            \centering 
            \includegraphics[width=\textwidth]{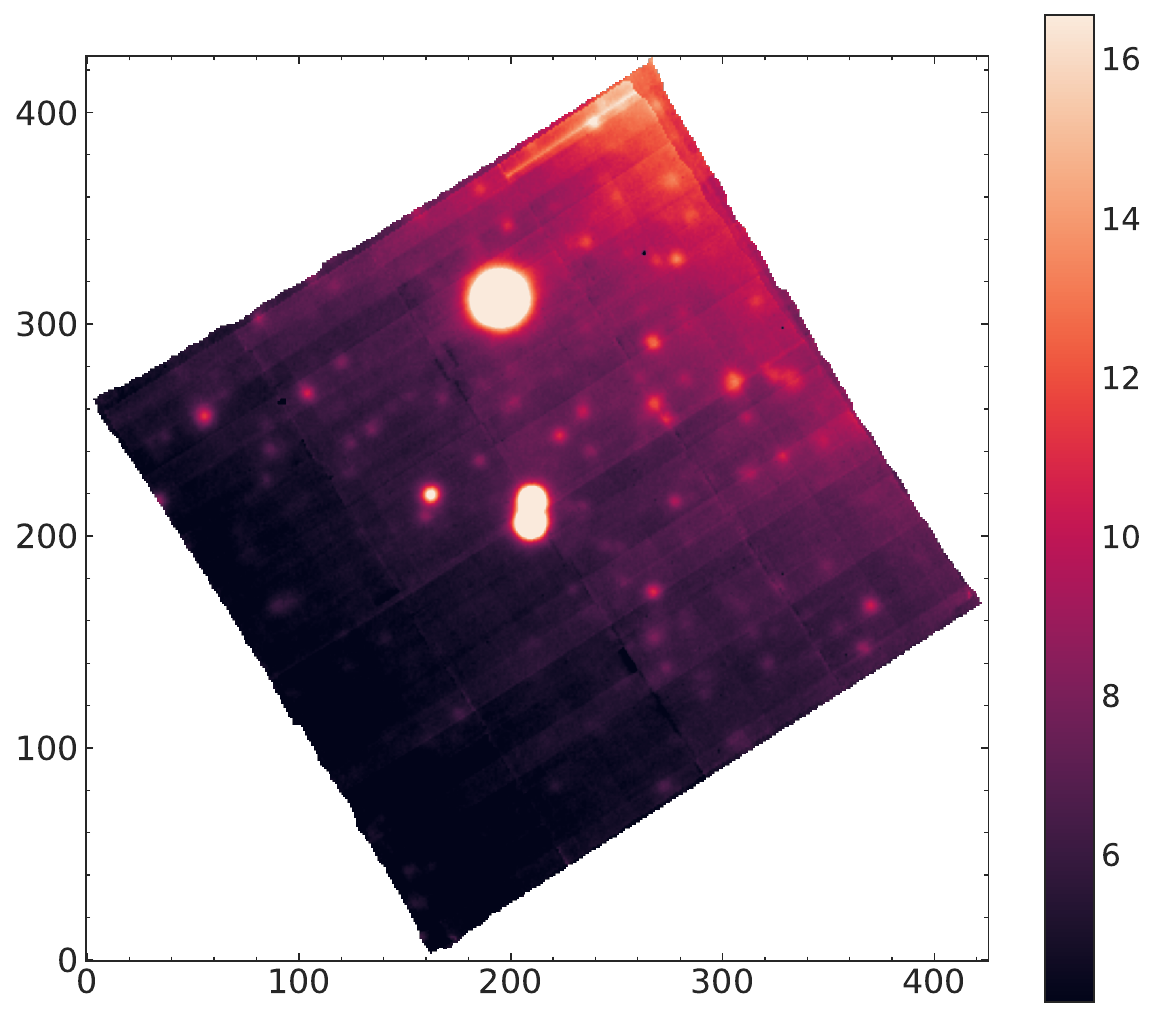}
            \caption{Field D}   
        \end{subfigure}
        \caption{S/N of MUSE data in the  $4800-7000 \,\AA$ wavelength range for all of the four fields around NGC~3311.} 
    \label{fig:fields_S/N}
        \label{fig:fields_S/N}
\end{figure*}

\begin{figure*}[t!]
    \centering
    \includegraphics[width=\textwidth]{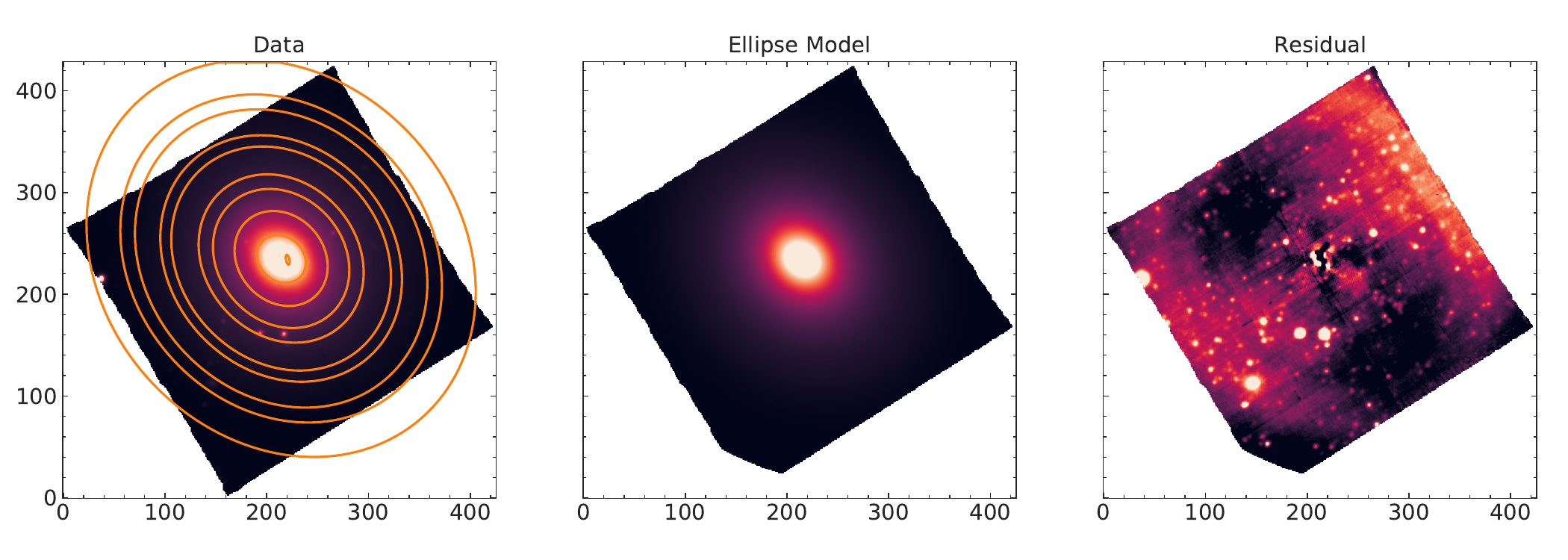}
    \caption{Isophotes for light of NGC~3311 in field A (left), resulting model 
    created to fit halo light of NGC~3311 (middle), and residuals (right); obtained by subtracting the first from the second. The residuals can be used to identify point-like sources in the central region.}
    \label{fig:fieldA_ellipse}
\end{figure*}

To identify sources, we used the python wrapper "source-extractor" from the sewpy\footnote{\url{https://sewpy.readthedocs.io/en/latest/}} module on the previously created S/N image of each field. During this procedure, we mask out regions containing bright foreground sources, such as the saturated star and its
diffraction spikes in fields B and C, as was also done in \citet{Barbosa2018}. Additionally, we used the python module "photutils.background" to remove the background signal from the data.
However, field A, which is the field containing the center of NGC~3311, is dominated by the very bright light of the galaxy itself, which has to be removed. To do this, we created a model elliptical galaxy from a list of isophotes using the python module "photutils.isophote" and determined the best fitting isophotes to the galaxy. We then subtracted this model from the original data, as can be seen in Figure~\ref{fig:fieldA_ellipse}. The isophotes, overplotted on the data, are shown in the leftmost panel, the model ellipse in the central one, while the rightmost image shows the residual, determined by subtracting the model from the data. Using this residual, we can find sources that are not part of the smooth light distribution of the galaxy, such as GCs, foreground stars, and background galaxies.

Through this procedure, we produced a catalog of sources for each field. In addition, using the sewpy python module, we determined some properties of each source during the extraction, such as the positions, ellipticities, (uncalibrated) magnitudes, and flux radii.

\subsection{Spectra extraction}

To extract the spectra of the identified sources from the MUSE datacubes, we used the astropy module "photutils.aperture". We tried different aperture sizes around each of the sources to determine the optimal aperture that yields the highest S/N.

\begin{figure}[t!]
    \centering
    \includegraphics[width=\columnwidth]{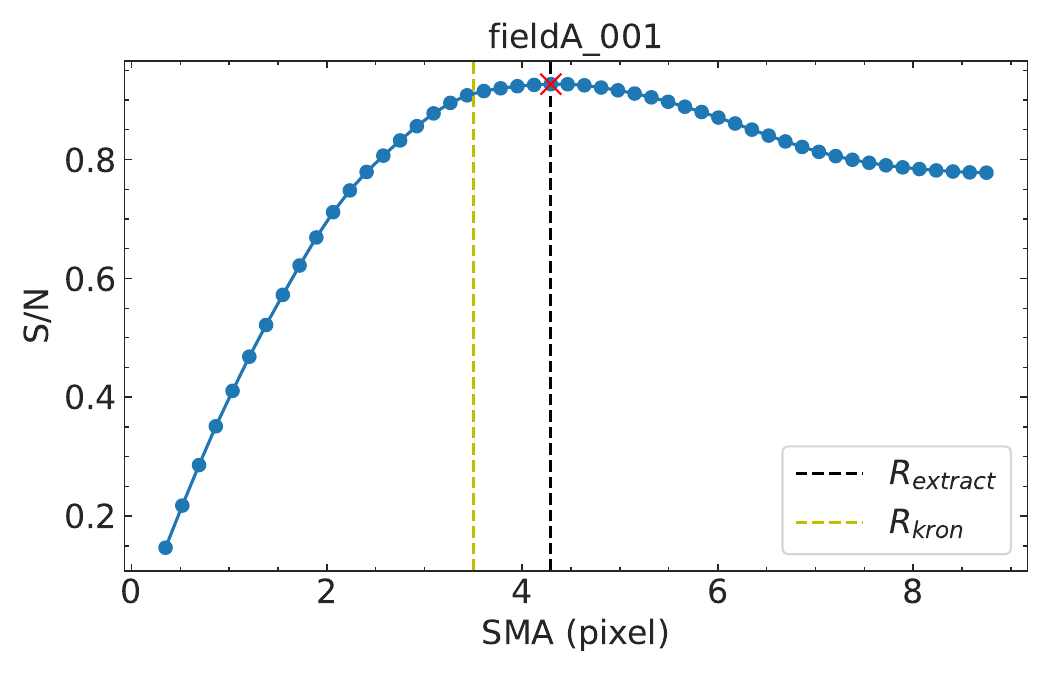}
    \caption{Resulting S/N as function of semi-major axis (SMA) of the elliptical aperture for a field A source. The optimal aperture (vertical dashed black line), at which we obtain the highest S/N, is used for the photometry.}
    \label{fig:apphot_maxS/N}
\end{figure}

Figure~\ref{fig:apphot_maxS/N} shows an example for a generic source in field A. The plot shows the semi-major axis (SMA) of the elliptical aperture plotted against the S/N. The vertical dotted yellow line shows the value for the Kron radius $R_{kron}$, which is a measure for the size of the object in pixels. The aperture that yields the maximum S/N, which is marked by the vertical dashed black line, is used for the photometry. This procedure is done for each of the sources individually. For non-defined cases, such as sources that have no prominent maximum or have more than one maximum, we used the median from all of the other aperture sizes in each field. The median is representative for the approximate size of a GC in each field. 
    
\begin{figure*}[t!]
        \centering
        \begin{subfigure}[b]{0.475\textwidth}
            \centering
            \includegraphics[width=\textwidth]{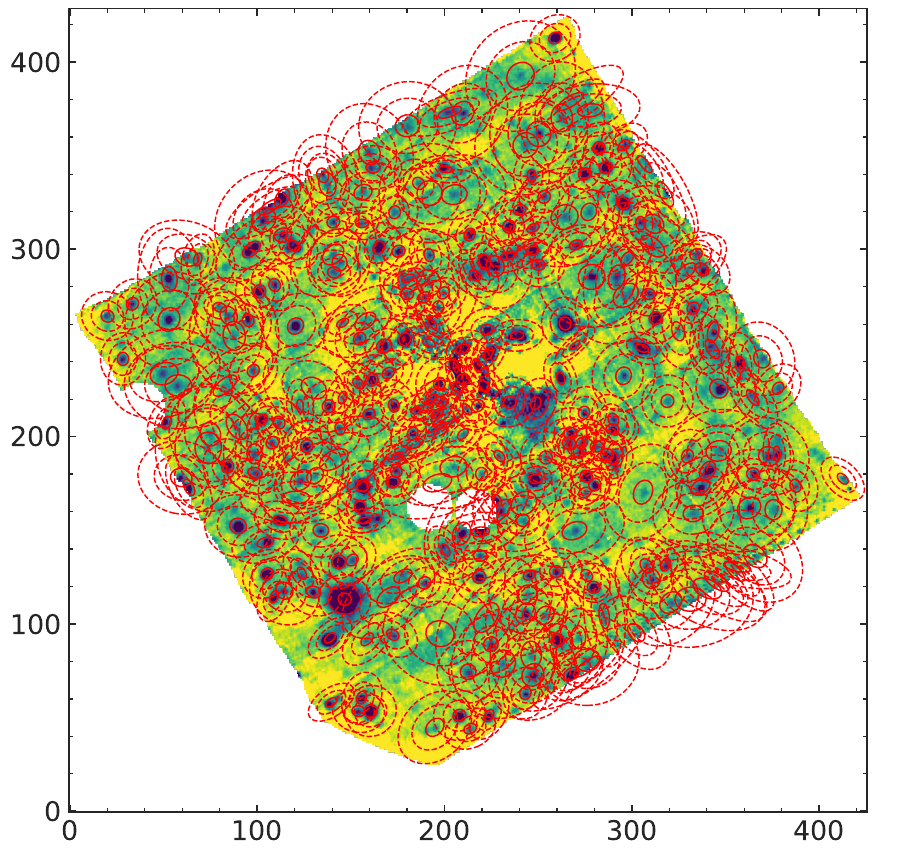}
            \caption{Field A}           \end{subfigure}
        \hfill
        \begin{subfigure}[b]{0.475\textwidth}  
            \centering 
            \includegraphics[width=\textwidth]{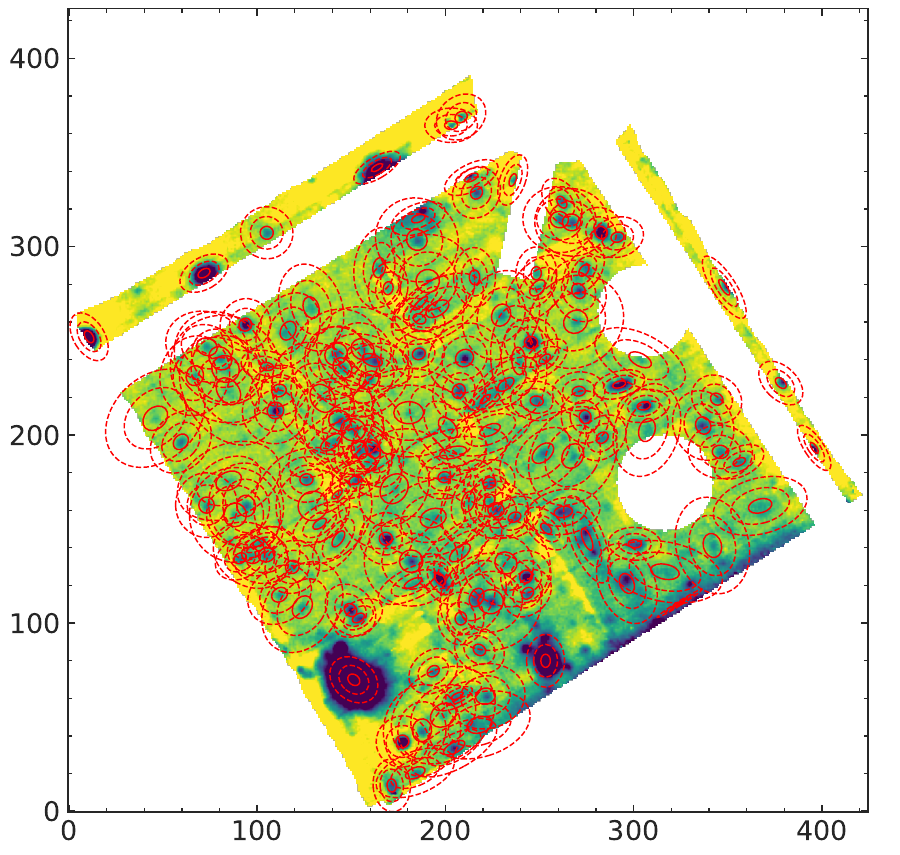}
            \caption{Field B}    
        \end{subfigure}
        \vskip\baselineskip
        \begin{subfigure}[b]{0.475\textwidth}   
            \centering 
            \includegraphics[width=\textwidth]{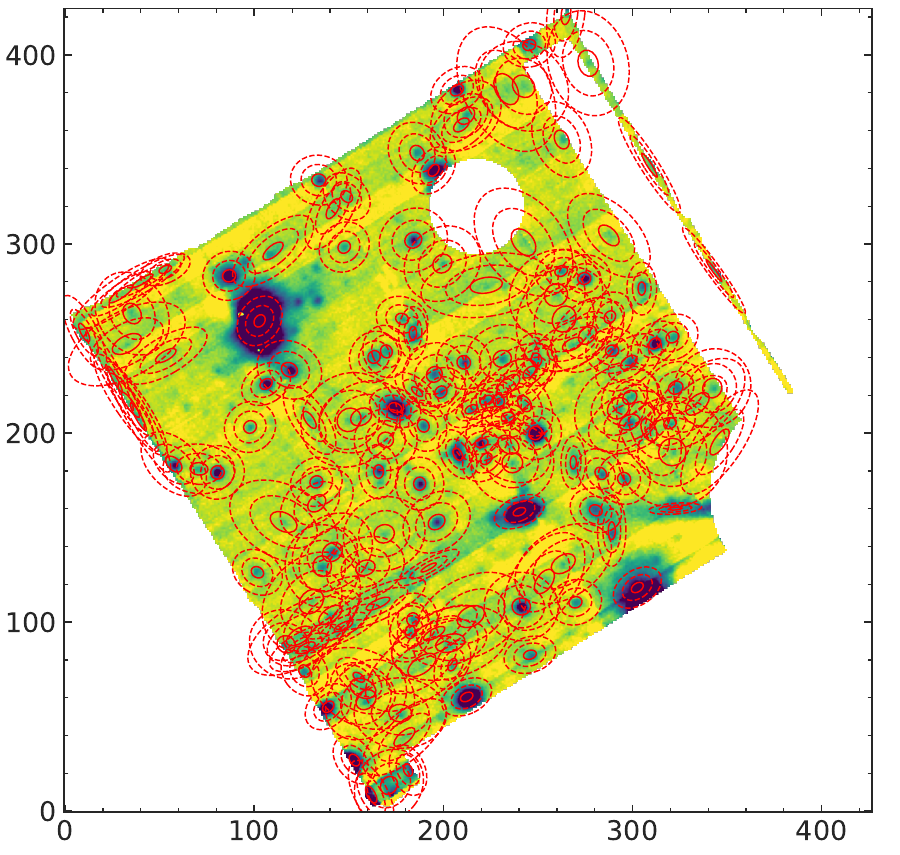}
            \caption{Field C}   
        \end{subfigure}
        \hfill
        \begin{subfigure}[b]{0.475\textwidth}   
            \centering 
            \includegraphics[width=\textwidth]{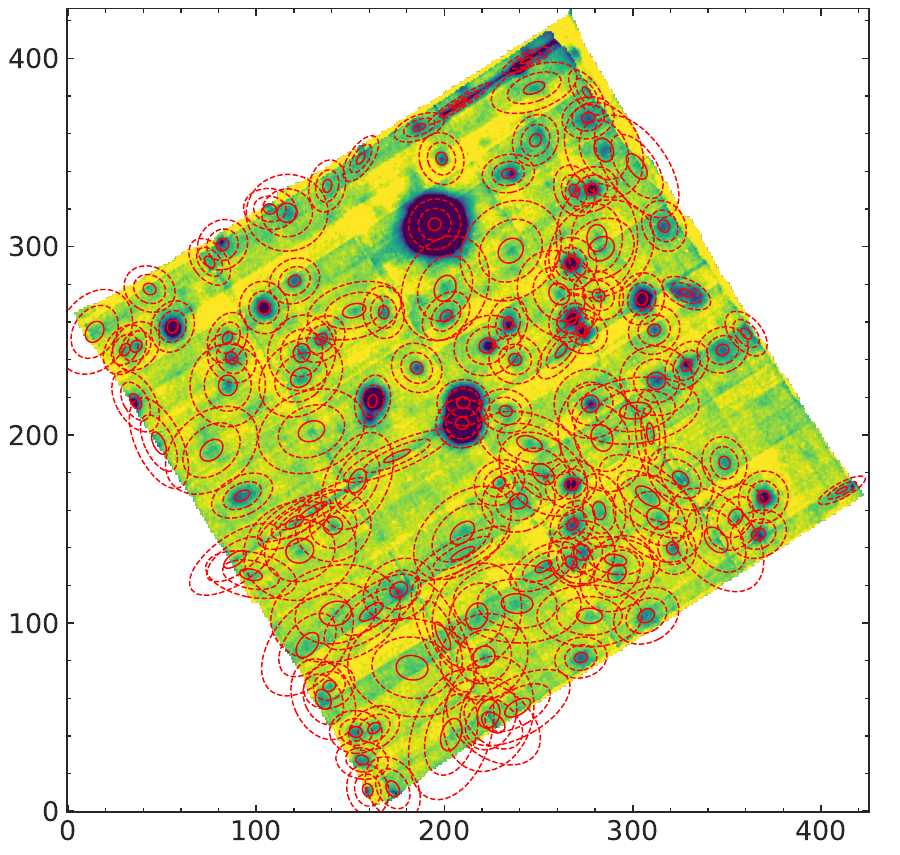}
            \caption{Field D}   
        \end{subfigure}
        \caption{Apertures and annuli used to extract highest S/N spectra, determined for the previously identified sources in all four fields. Darker regions show higher signals, i.e., brighter sources.} 
        \label{fig:fields_apphot}
    \end{figure*}

We used individually determined elliptical annuli around each source to remove any signals and contamination from the background and hence obtained clean spectra of these candidate GCs. Each annulus has an inner radius of $2.5\,R_{kron}$ from its target source. This relatively large value was chosen to avoid including any contribution from the source to the background that we then subtracted. We chose an outer radius of $4\,R_{kron}$ for the annuli to allow for a large enough area to collect a meaningful background signal. For the annuli that overlap with other sources, we masked out those 
sources to avoid their contaminating effects. Figure~\ref{fig:fields_apphot} shows the apertures and annuli for all sources in all of the fields. We note that the bright foreground stars have been masked out. We acknowledge that there are a few cases in which the chosen anulii are too small and intercept the source itself. However, we stress that we are interested in extracting GCs, which are point-like sources. Hence, a suboptimal treatment of the extended sources is acceptable as it will not affect the results presented in the paper.

At this point, for each of the identified sources,
we obtain the spectrum within the optimal aperture and the flux within the annulus for the entire MUSE wavelength range of $4650-9300 \AA$. In order to obtain a sky-subtracted and clean spectrum of each source, we remove the background signals by subtracting the flux within the annulus from the flux within the aperture. The total amount of extracted sources in each of the four fields is listed in the second column of Table~\ref{tab:Nsources} ("sources").
    
\subsection{Selection of bona fide GC candidates} \label{sec:bonafide}
Having extracted all the spectra of the sources in the surroundings of NGC~3311, the next step is to identify which are the most reliable, bona fide GC candidates. In fact, the objects found by the source extractor could be background galaxies, dwarf galaxies, or GCs. 
In particular, first of all, we removed flux outliers using a sigma clipping routine, keeping only sources with a magnitude within five standard deviations from the mean.
Secondly, we applied 
three independent selection criteria to narrow down the number of GC candidates according to their morphology, radial velocities, and spectral characteristics. These quantitative selection criteria are described below.

\subsubsection{Morphologies} \label{subsubsec:morph}

Since GCs are spherical, we only considered sources that have an ellipticity $\epsilon\leq0.3$ to be GC candidates. Finally, we discarded extended sources by excluding objects with radii larger than the seeing in each field (0.92, 1.4, 1.3, and 1.8 arcseconds, for fields A, B, C, and D, respectively). We are aware that the seeing conditions of the four fields were quite different during the observations and that calibrating the morphological selection on field A might cause us to miss possible candidates in other fields. However, since field A is the densest field and the majority of the selected sources are in field A, we decided to implement the most conservative morphological criteria on the selected sources

\begin{figure*}[t!]
        \centering
        \begin{subfigure}[b]{0.475\textwidth}
            \centering
            \includegraphics[width=\textwidth]{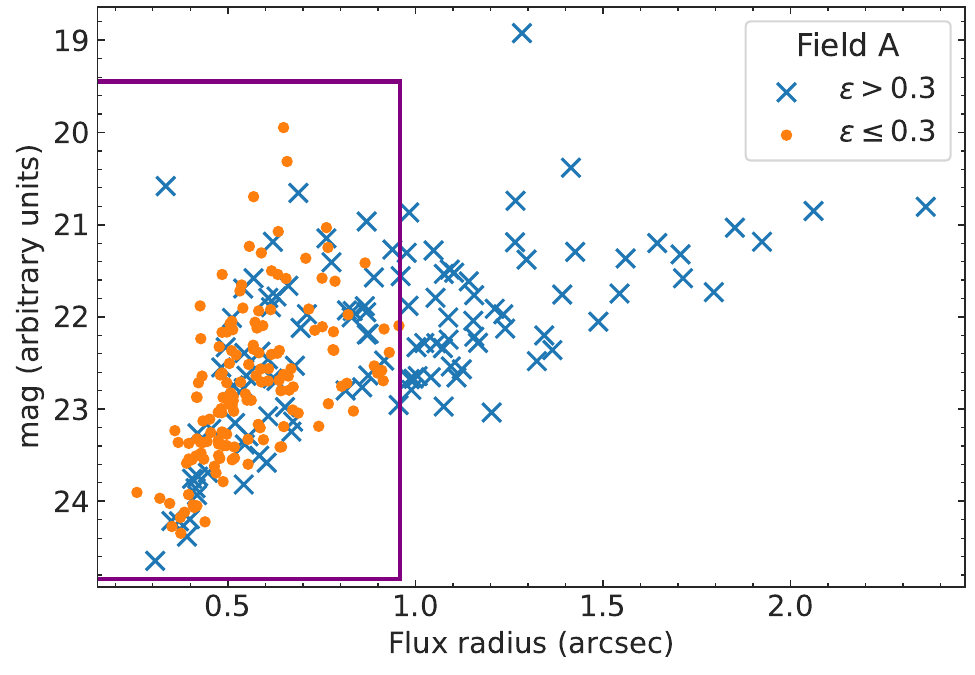}
            \end{subfigure}
        \hfill
        \begin{subfigure}[b]{0.475\textwidth}  
            \centering 
            \includegraphics[width=\textwidth]{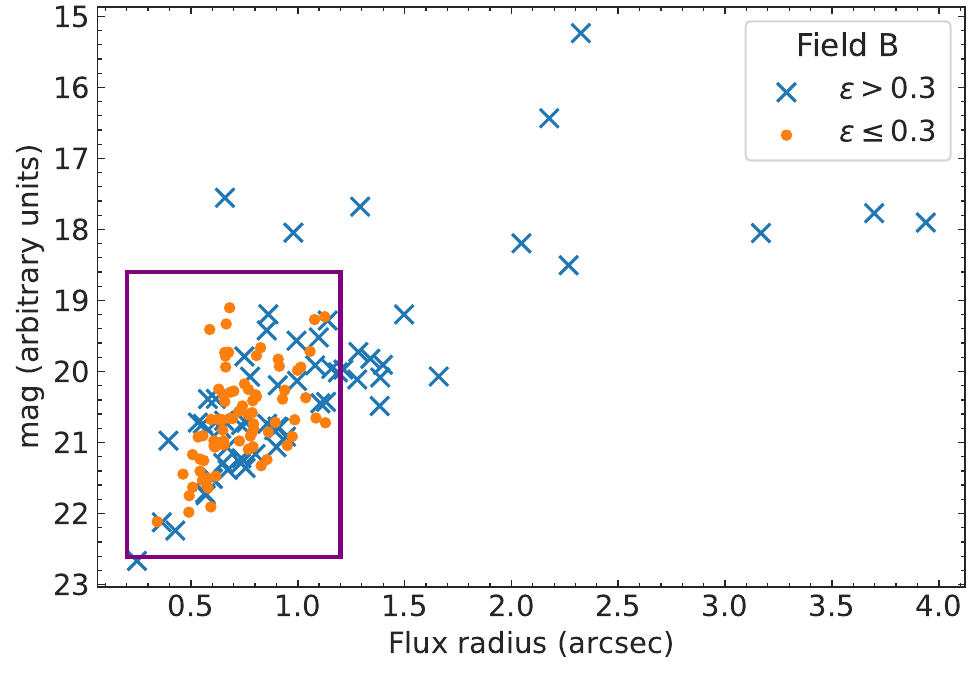}
        \end{subfigure}
        \vskip\baselineskip
        \begin{subfigure}[b]{0.475\textwidth}   
            \centering 
            \includegraphics[width=\textwidth]{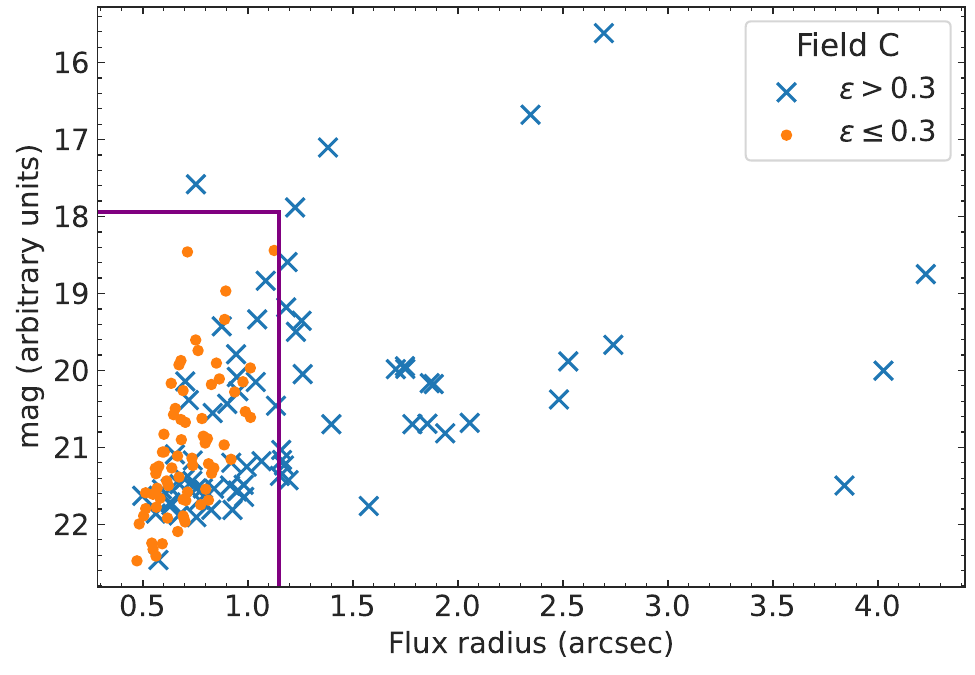}
        \end{subfigure}
        \hfill
        \begin{subfigure}[b]{0.475\textwidth}   
            \centering 
            \includegraphics[width=\textwidth]{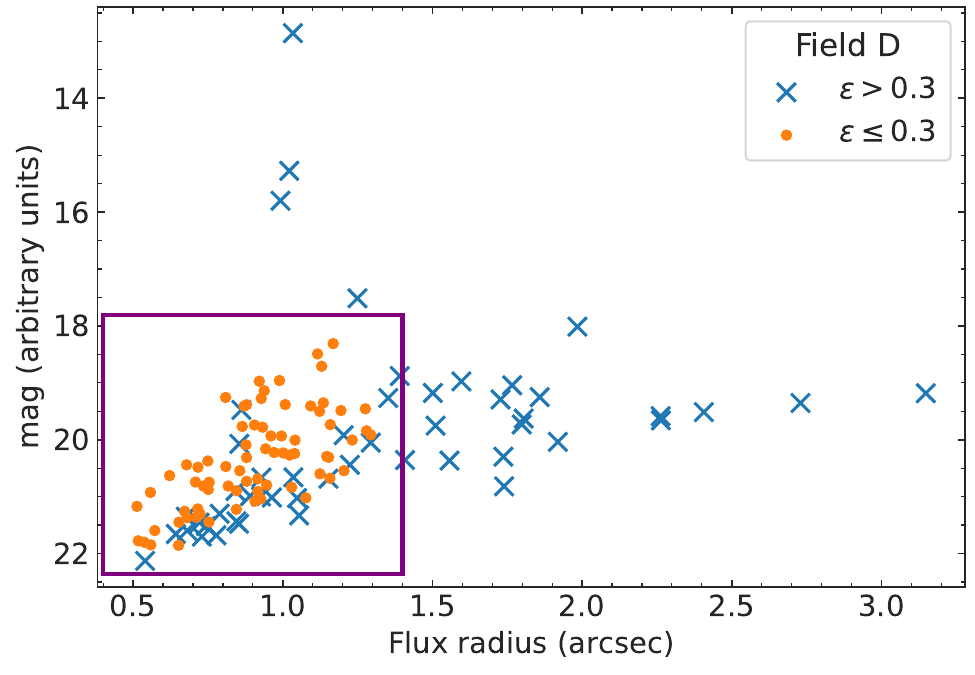}
        \end{subfigure}
        \caption{Magnitude of extracted sources in the four fields in relation to their flux radius. Sources with eccentricities $\epsilon\leq0.3$ are shown as orange dots, while sources with higher eccentricities are shown as blue crosses. Purple boxes show the accepted magnitude values (5 standard deviations from the mean). The sources with $\epsilon\leq0.3$ within the purple box in each image are considered GC candidates.}        \label{fig:gc_candidates}
    \end{figure*}

Figure~\ref{fig:gc_candidates} shows the (uncalibrated) magnitude of the extracted sources of all fields in relation to their flux radius. Sources with eccentricities $\epsilon\leq0.3$ are represented by orange dots, while sources with higher eccentricities are shown as blue crosses. Purple boxes indicate our selection criteria. Only sources with $\epsilon\leq0.3$ (orange dots) that lie within the purple box, defined by the seeing in each field, are considered GC candidates. The number of objects that pass this morphological selection are listed for each of the four fields in the third column of Table~\ref{tab:Nsources} (morphology). They represent roughly $50\% $ of the total of the extracted sources. 

\subsubsection{Radial velocities}  \label{subsubsec:radvel}

Since NGC~3311 has a velocity of around $3800$ km/s, the GCs that are gravitationally bound to this galaxy would have compatible velocities. We considered all sources in the velocity range $2500-5000$ km/s as bound to the system. To measure the radial velocity on each individual spectrum that passed the morphology selection criteria, we used the function "template\_correlate" from the astropy module "specutils.analysis". This function computes a cross-correlation of the observed input spectra and some template spectra, and it determines the difference in velocity.

\begin{figure}[t!]
    \centering
    \includegraphics[width=\columnwidth]{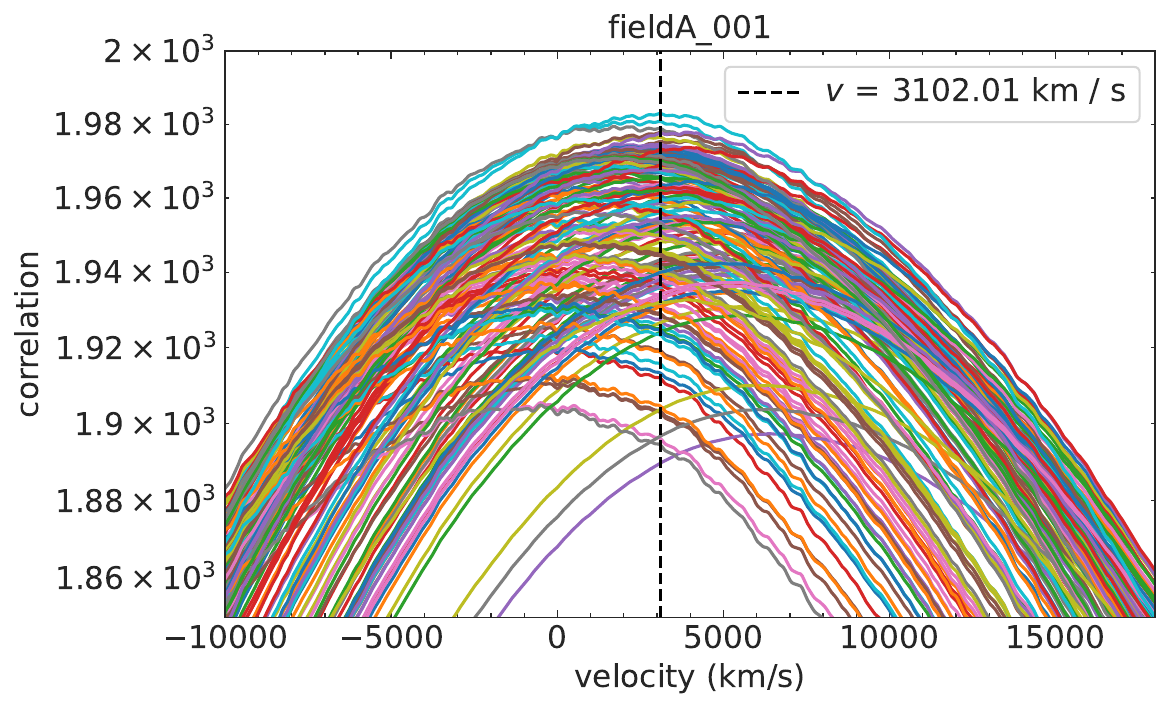}
    \caption{The correlation between the observed source spectrum and the MILES stellar templates for a random field A source. The most likely velocity, shown by the dashed vertical line, is the one with the highest correlation.}
    \label{fig:rv_corr}
\end{figure}

As the template spectra, we used the entire MILES \citep{MILES2010} empirical stellar spectral library that comes with {\tt pPXF} \citep{PPXF2017}. The most likely velocity is the one with the highest correlation. During the process of correlating the spectra, we also removed sources that have a negative median flux and did not consider them in our further calculations. Figure~\ref{fig:rv_corr} shows an example of a field A source with the velocity plotted against the correlation with all of the 150 MILES stellar spectra. There are also cases where the correlation curve has a second peak. In such cases, we could not be absolutely sure if the highest peak was really the correct one, but we used it as our best estimate since it had the strongest correlation signal.

\begin{figure}[t!]
    \centering
    \includegraphics[width=\columnwidth]{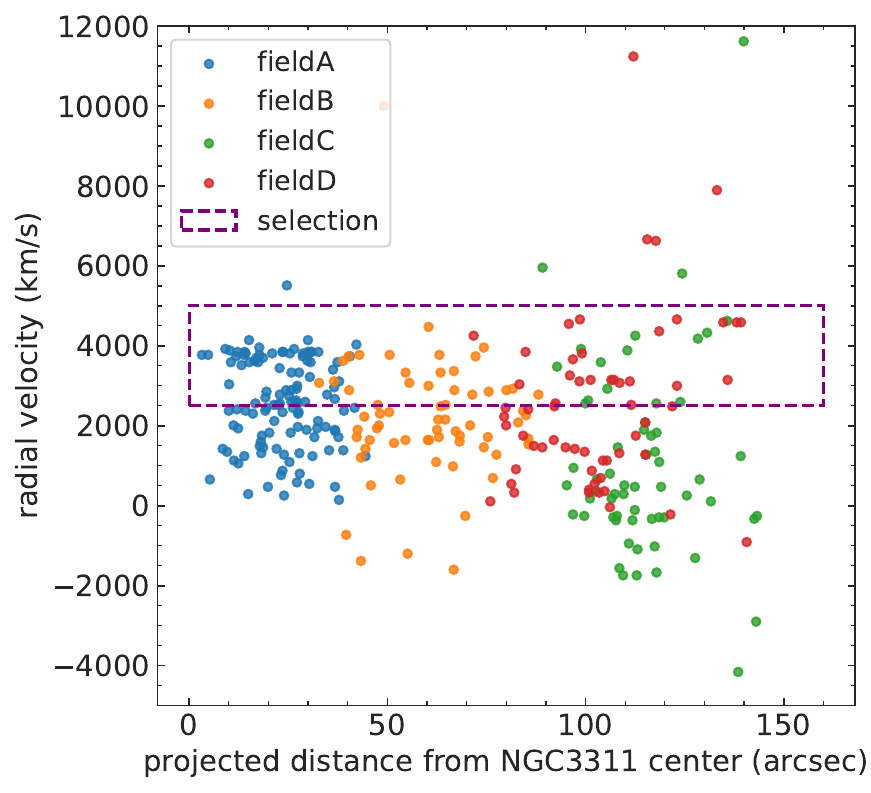}
    \caption{Projected distance of GC candidates (Sect.~\ref{subsubsec:morph}) from the center of NGC~3311 versus their radial velocities. The different fields are shown in different colors, as indicated in the legend. The purple box includes all sources within our selected velocity range of $2500-5000$ km/s.}
    \label{fig:vel_dist}
\end{figure}

Figure~\ref{fig:vel_dist} shows the phase diagram of the projected distance of the sources from the center of NGC~3311 against their radial velocities. In this plot, we only include GC candidates that fulfill the morphology criteria (see Section~\ref{subsubsec:morph}) and were not removed due to a negative median flux. The different fields are shown in different colors, as indicated in the legend. The purple box represents our selected velocity range of $2500-5000$ km/s. 
The number of objects that pass the velocity criterion are listed in the fourth column of Table~\ref{tab:Nsources} (velocity) for each of the four fields.

We note that the velocity dispersion of the galaxy is approximately $\sigma \gtrsim 200$ km/s, meaning that our GC candidates are within a range of $5-6\sigma$ from the systemic velocity. However, in our case, the larger range in velocity is justified by NGC~3311 being the BCG of the Hydra I cluster, with the Hydra I cluster galaxies having a velocity dispersion of $\sigma_{cluster} = 700 $\,km/s \citep{Christlein2003}. Applying a more conservative selection within $3\sigma$ of the systemic velocity would recover approximately 2/3 of our final GC candidates, showing that the majority of our selected GC candidates are indeed within a velocity range (approximately $\pm 3\sigma$) consistent with the central velocity dispersion of NGC~3311. It can also be seen in Figure~\ref{fig:vel_dist} that the field A sources, which constitute most of our GC candidates, align quite well with the velocity of the galaxy NGC3311.

\subsubsection{Emission lines}
The individual spectra that passed the previous criteria were checked for the presence of emission lines to assess whether the selected candidates are bona fide GCs. In fact, since the stellar populations of GCs are expected to be old, their early-type stars are long dead and their spectra are made up of mostly K- and M-type stars, which have prominent  calcium, sodium, and magnesium stellar absorption features and show no emission.

Since the S/N of individual spectra might be too low to precisely identify absorption features, we focused on whether the spectra of GC candidates have emission lines, since these are easier to identify. Sources with emission lines are not GCs, but most likely background galaxies. By examining the individual spectra, we were able to identify a sizeable number of sources with emission lines, particularly in field D, which turned out to have only emission line sources. We note that, at this stage, we only removed sources with strong and evident emission lines.\footnote{Given the very low S/N of single spectra, it is not always possible to differentiate between weak emissions and peaks in the noise. This was further investigated in the stacked spectra.}  

\section{Final GC catalog}
\label{sec:catalog}

We list the number of bona fide GC candidates as those selected from the morphological, velocity, and no-emission criteria in the rightmost column of Table~\ref{tab:Nsources}. 
Initially, we started from a total of 680 sources, half of which passed the morphological and photometrical criteria. 
Then, from an intermediate sample of 369 unresolved point-like and round sources, only $49$ candidates satisfied both the velocity and the absence of emission line criteria. Most of the bona fide GCs are found in field A. These sources selected from three of the four MUSE cubes have the highest probability of  being GCs (since they are unresolved) within the velocity interval of the BCG NGC~3311, and they display no emission lines. 
In the rest of our work, we only considered these 49 GC spectra.

As described in Section~\ref{sec:bonafide}, flux outliers were removed by applying a sigma clipping selection, which can lead to potential sources being excluded. To test whether this has an effect on our final GC catalog, we also ran the candidate selection without applying this filter. In this case, we obtained the exact same sources as with the filter, except in field D, which yielded three additional sources. Cross-correlating their spectra with the MILES spectral library, as we did with the sources, revealed radial velocities for these sources that are outside of our selection range for bona fide GCs. Since these three sources would be excluded from the further analysis steps anyway, the sigma clipping does not influence our final result.

\begin{figure}[t!]
    \centering
    \includegraphics[width=\columnwidth]{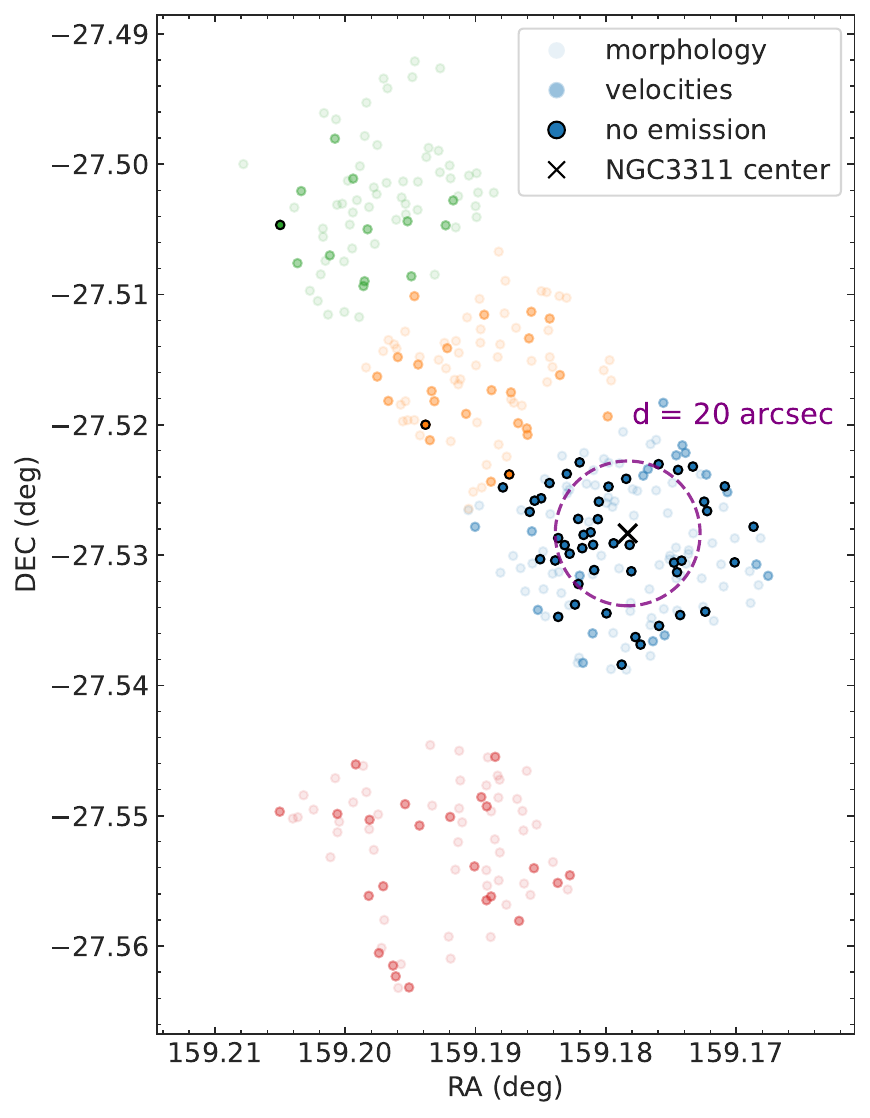}
    \caption{Sky coordinates of all extracted sources that fulfill the morphology criteria. The most transparent sources fulfill only the morphology criteria (Sect.~\ref{subsubsec:morph}). The sources with medium transparency additionally fulfill the velocity criteria (Sect.~\ref{subsubsec:radvel}). Finally, the sources with black contours are those that fulfill all previous criteria and have no emission lines, and are therefore the final GC candidates. The colors of the fields are the same as in Figure~\ref{fig:vel_dist}, and the center of NGC~3311 is marked by a black cross. The purple circle marks a projected distance of 20\,arcsec from the galaxy center.}
    \label{fig:coords}
\end{figure}

Figure~\ref{fig:coords} shows the sky coordinates of the GC candidates, which were calculated using the world coordinate system (WCS) astropy module,  which  converts the pixel positions into sky positions. The sources with the highest transparency are those that fulfill only the morphology criteria described in Section~\ref{subsubsec:morph}. The sources with the medium transparency are those that additionally fulfill the velocity criteria defined in Section~\ref{subsubsec:radvel}.  
Finally, the least transparent sources with a black contour are those that fulfill all previous criteria and exhibit no emission lines. These are the final 49 bona fide GC candidates. The colors of the fields are the same as in Figure~\ref{fig:vel_dist}, and the center of NGC~3311 is marked by a black cross.

\begin{table}[t!]
    \centering
    \caption{Number of GC candidates considering different criteria.}
    \begin{tabular}{ccccc}
    \hline
    \hline
        field & sources & morphology & velocity &  no emission \\
         \hline
        A & 268 & 147 & 64 & 46 \\ 
        B & 143 & 79 & 24 & 2\\
        C & 143 & 70 & 13 & 1\\
        D & 126 & 73 & 23 & 0\\
        \hdashline[0.5pt/1pt]
        all & 680 & 369 & 124 & 49\\
        \hline
    \end{tabular}
            \begin{tablenotes}
      \item \textbf{Notes.} Number of sources extracted in each field with "source-extractor", the number of candidate GCs left after applying the morphology ($\epsilon \le 0.3$) and velocity ($v_r\in [2500,5000]$ km/s) criteria, and the number of sources left when further discarding objects showing emission lines in their spectra.
    \end{tablenotes}
    \label{tab:Nsources}
\end{table}

\subsection{Color distribution of bona fide GCs in NGC~3311}\label{subsec:colors}

The analysis of the colors of GCs can provide information on their properties and also on the accretion history of the host galaxy. Often, a bimodal distribution is found for the GC colors (e.g., \citealt{Beasley2018}), which exhibit a blue and a red  peak. The origin of this double-peaked distribution can be linked with distinct GC populations of different ages and/or metallicities, which should emerge in the two-phase formation scenario \citep{Oser2010,Hilz2012}.  Typically, the redder GCs are formed in situ with the same metal-rich stellar component of the host galaxy, while the bluer ones have been accreted from more metal-poor satellites \citep{Beasley2018}. When this is the case, the two GC populations also display different kinematical properties, with the red GCs more closely following the stellar kinematics and the blue GCs having  less relaxed kinematics \citep{Schuberth2010, Coccato2013, Napolitano2014, Pota2018}. 

If a galaxy underwent an extended mass accretion, the redder, older, more metal-rich GCs, which formed in situ with the bulk of the host stars, are expected to be more centrally concentrated. The bluer, more metal-poor and/or younger, GCs are instead expected to be distributed more in the outskirts, as the least massive satellites are disrupted at larger radii \citep{Amorisco2017}. To verify whether such spatial segregation is present in NGC~3311, we split the GCs into two subsamples depending on their projected distance $d$ to the galaxy center. We thus divided them into an inner ($d<20\,$arcsec) and an outer ($d>20\,$arcsec) GC subsample, as shown by the purple circle in Figure~\ref{fig:coords}. 
The chosen dividing radius ($d=20\,$arcsec $\sim4$ kpc) on one hand allowed us to have a similar number of GCs in the inner and outer subsample, but on the other hand it is also motivated by the results obtained in \citet{Barbosa2021}. In particular, from their Figures~~5 and 7, it appears that the light in the innermost $\sim$3-4 kpc is dominated by the photometric components A and B identified in \citep{Barbosa2018}. 
These components are both characterized by stars with a velocity dispersion of $\sigma\sim 200$ km/s, old ages, and solar or slightly solar metallicity. At galactocentric distances larger than $4$ kpc, both the stellar velocity dispersion $\sigma$ and the [$\alpha$/Fe] abundance ratio rise to $\sigma > 300$ km/s and $\simeq 0.2,$ respectively, while the stellar age starts to decrease down to $\simeq 8$ Gyr.

Even though the 49 GC candidates are preferentially found in the central region, we examined the color distribution of the GCs in NGC~3311, divided into an inner and outer subsample. 
Using the astropy package speclite, we determined the SDSS $r$ and $i$ magnitudes, since the range of those two filters does overlap with the wavelength range of the MUSE  spectra. We hence determined the $r-i$ color of the two GC populations, and we plot their distribution in Figure~\ref{fig:sdss_r-i}.
While the distribution is broad and does not display an evident bimodal distribution for the GC colors, there is a clear correlation. The outer GCs, in fact, cluster  toward the bluer edge of the distribution, while the inner ones cluster toward the redder edge, as expected from the two-phase formation scenario. Similar color gradients with a radius for GC stellar populations in ETGs were also found in other studies \citep{Tortora2010, Li2018, Forbes2018}.



\subsection{Comparison with previously published color distribution for the NGC~3311 GCs }\label{subsec:colorsHempel}

The GC population in NGC~3311 was studied by \citet{Hempel05} using combined optical and near-infrared (NIR) photometry. The optical data were obtained with the F814W and F555W filters on HST+WFPC2, while the NIR data came from HST+NICMOS through the F160W filter (which is approximately an $H$ -band filter). In particular, the color magnitude diagram $V$ versus $V-I$ and the $(V-I)$ versus $(V-H)$ color-color diagram for unresolved sources in NGC~3311 were analyzed in the paper. These diagrams returned broad distributions with $\Delta (V-I) \simeq 0.6$ mag and $\Delta (V-H) \simeq 1.8$ mag. Once these broad distributions were compared with single stellar population (SSP) isochrones, those with $(V-H) > 2.5$ and $(V-I)< 1.1$ required ages of 5 Gyr or less \citep{Hempel05}. 

\begin{figure}[t!]
    \centering
    \includegraphics[width=\columnwidth]{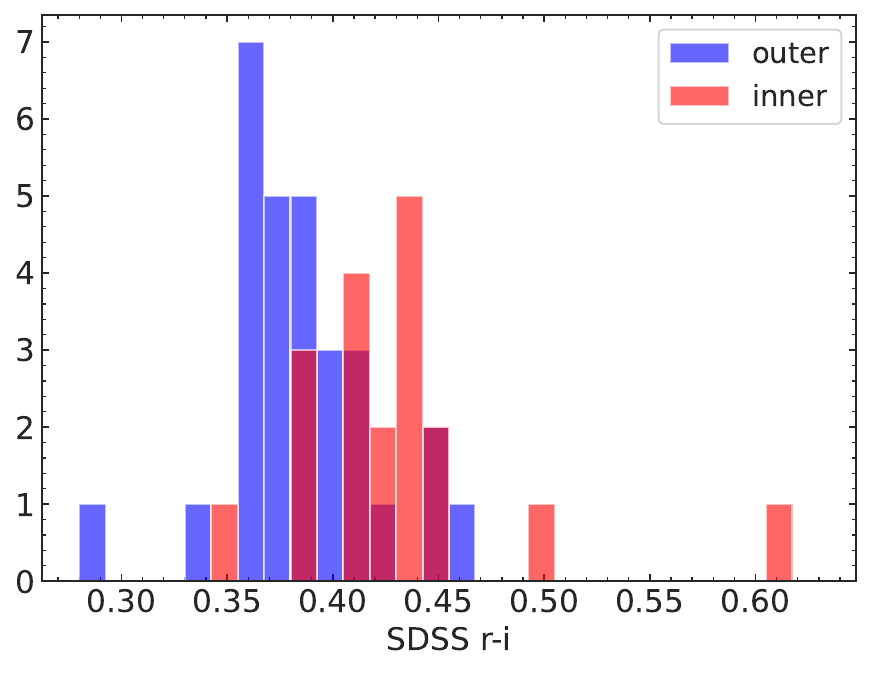}
    \caption{SDSS $r-i$ color distribution of inner ($d<20\,$arcsec, red) and outer ($d>20\,$arcsec, blue) GCs, as determined with the astropy package speclite.}
    \label{fig:sdss_r-i}
\end{figure}

In Section~\ref{subsubsec:radvel}, we argued for the presence  of emission lines in a significant fraction of unresolved sources, within a  suitable radial velocity range to be bound to NGC~3311. One can thus conclude that the contamination by background emission line objects might be the root cause of the claim of an intermediate age population of unresolved sources in \citet{Hempel05}. To prove this statement, in Section~\ref{sec:stelpop}, we  investigate the stellar population properties of stacked spectra of the bona fide GCs identified in this work. Since \cite{Hempel05} did not publish the list of their GC candidates, we were unable to match them with ours. Performing a cross-match of our GC candidates with the HST catalog available from the HST archive\footnote{\url{https://mast.stsci.edu/portal/Mashup/Clients/Mast/Portal.html}} with a matching radius of 1 arcsec yields 27 matches, of which, unfortunately, only three are inner GCs that have both a F555W and F814W magnitude. Therefore, we are unable to determine if the HST color distribution also exhibits a similar bimodality to that in Figure~\ref{fig:sdss_r-i}.

\section{Stacked spectra of bona fide GCs in NGC~3311}
\label{sec:stacking}

\begin{figure}[t!]
        \centering
        \begin{subfigure}[b]{0.5\textwidth}
            \centering
            \includegraphics[width=\textwidth]{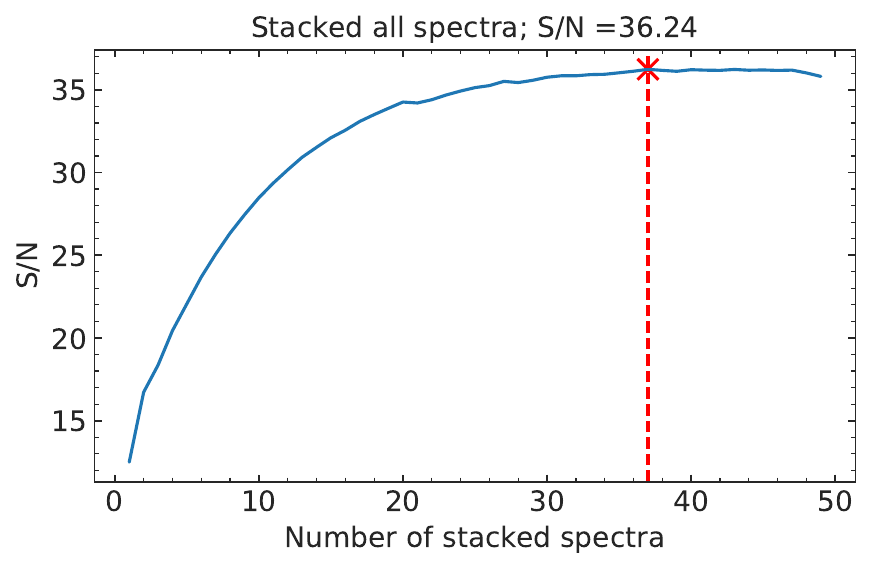}
            \end{subfigure}
        \hfill
        \begin{subfigure}[b]{0.5\textwidth}  
            \centering 
            \includegraphics[width=\textwidth]{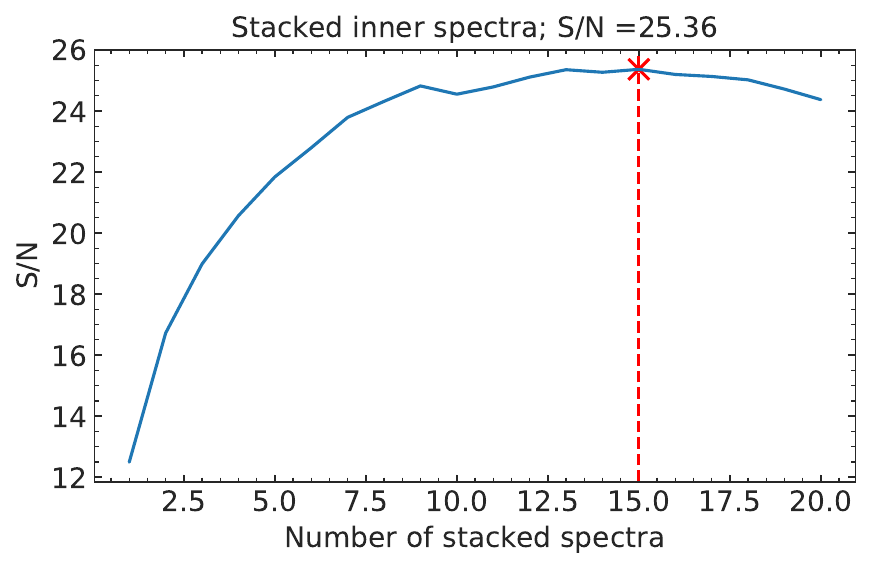}
        \end{subfigure}
        \vskip\baselineskip
        \begin{subfigure}[b]{0.5\textwidth}   
            \centering 
            \includegraphics[width=\textwidth]{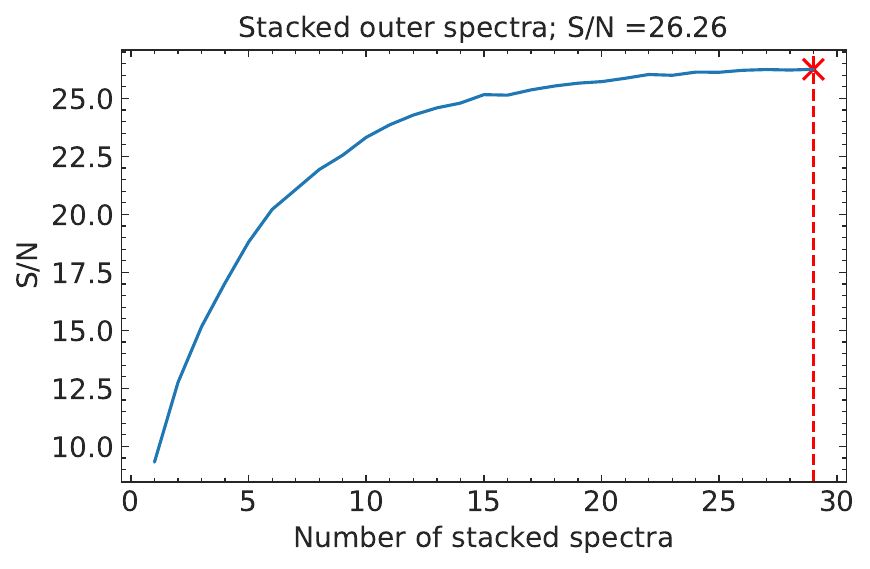}
        \end{subfigure}
        \caption{S/N of stacked spectra of all (top), inner (middle), and outer (bottom) GC populations, depending on the number of GC spectra that are stacked. The individual GC spectra have been sorted according to their S/N and stacked beginning with the highest S/N.} 
        \label{fig:GCstacksSN}
    \end{figure}

\begin{figure*}[t!]
        \centering
        \begin{subfigure}[b]{\textwidth}
            \centering
            \includegraphics[width=\textwidth]{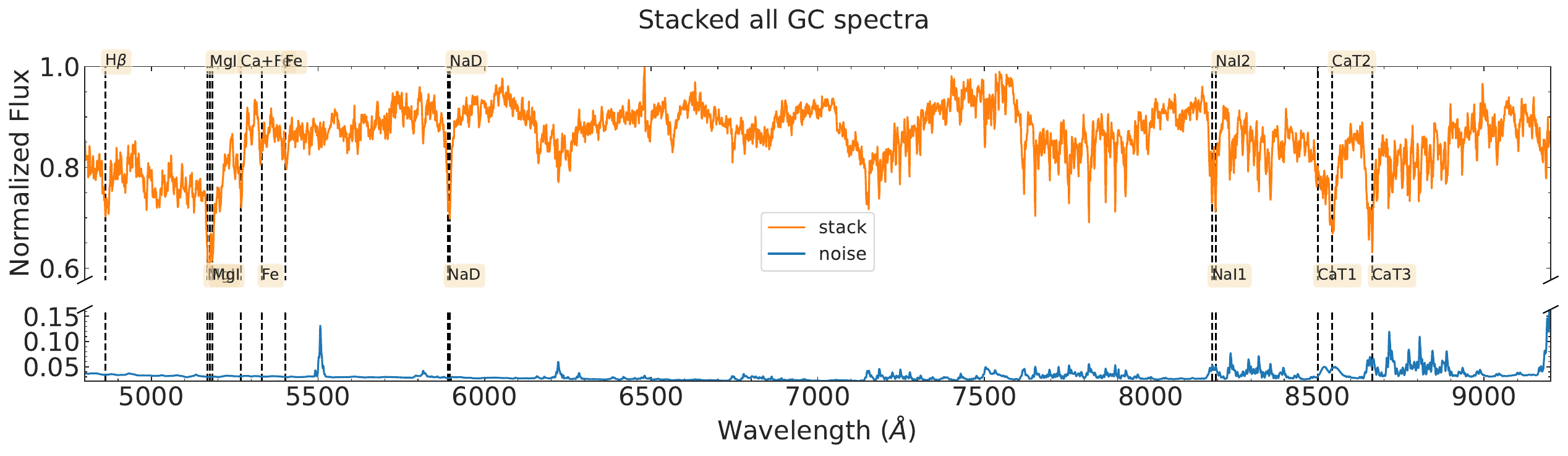}
            \label{fig:GCstack_all}
            \end{subfigure}
        \hfill
        \begin{subfigure}[b]{\textwidth}  
            \centering 
            \includegraphics[width=\textwidth]{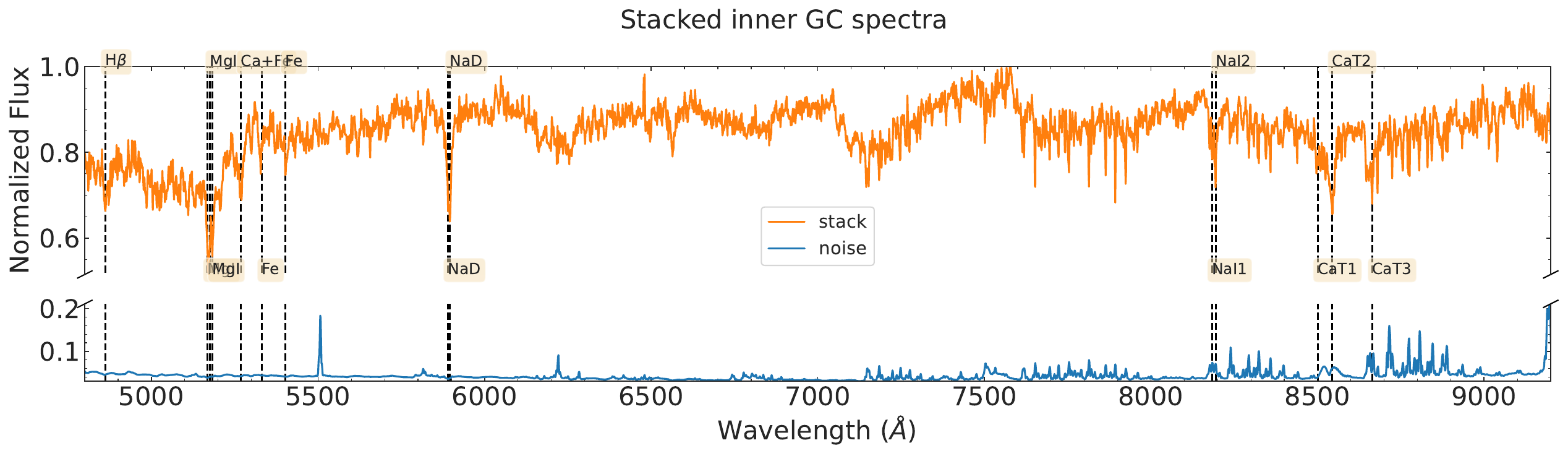}
            \label{fig:GCstack_inner}
        \end{subfigure}
        \vskip\baselineskip
        \begin{subfigure}[b]{\textwidth}   
            \centering 
            \includegraphics[width=\textwidth]{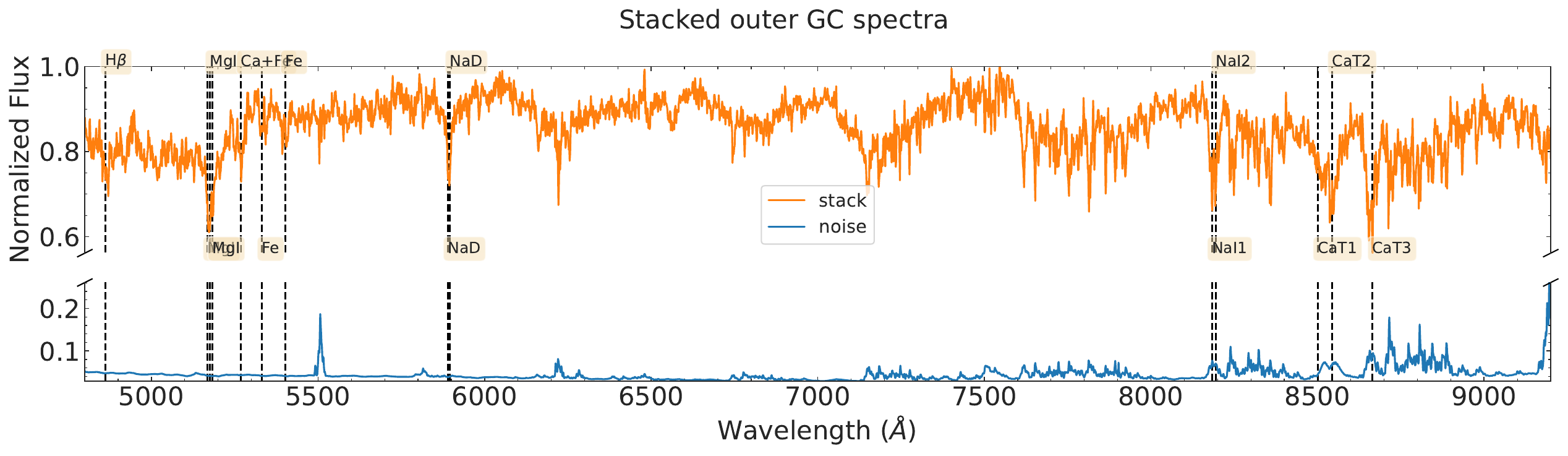}
            \label{fig:GCstack_outer}
        \end{subfigure}
        \caption{Stacked spectra of all,  inner, and outer GCs, along with the positions of some prominent absorption lines (MgI, Ca, Fe, NaD, NaI, CaT).} 
        \label{fig:GCstacksspectra}
    \end{figure*}    
    

In order to be able to reliably constrain the stellar population age, metallicity, and [$\alpha$/Fe] from the optical spectra, it is necessary to have a S/N of at least 10-15 per dispersive element  \citep{Costantin2019}. Unfortunately, the individual spectra of the 49 GC candidates do not reach this threshold. Their S/N values cover a range from S/N 1 to 12, which is not sufficient to characterize the most prominent absorption features in the optical.  
Thus, to study the stellar populations of the GCs in NGC~3311, it is necessary to stack spectra of a suitable number of bona fide candidates. Although stacking does indeed blur individual differences of the GC spectra, it makes it possible to characterize the GC population properties. The quantities we derive are averaged over the luminosities of the selected candidates. While this has obvious limitations, it provides a first insight into the stellar population of the GCs of this BCG.

Before stacking, we shifted all the spectra to the rest frame. We first used the radial velocities of the sources determined in Section~\ref{subsubsec:radvel}. However, when looking at the individual spectra of the GC candidates, plotted along with the locations of some prominent absorption lines (MgI, Ca, Fe, and NaD), we realized that not all of the radial velocities determined in Section~\ref{subsubsec:radvel} were accurate. Although about half of the spectra were misaligned with the rest frame, most of them only required a small additional shift. Since the number of spectra to inspect is manageable, we visually inspected them by eye, each time manually determining the required additional shift in velocity. 

At this point, we summed up the spectra to increase the final S/N. To perform an optimal stacking procedure, we sorted the spectra according to S/N and stacked them starting with the highest S/N spectrum. After a new spectrum was added to the total stacked spectrum, we measured the S/N of the resulting stack and only continued if the new S/N was higher than the previous one. In this way, we were able to determine the amount of spectra to be stacked to obtain the highest possible S/N spectrum. This is illustrated in the top panel of Figure~\ref{fig:GCstacksSN}. 

Moreover, using the same procedure, we also built two additional stacked spectra for the inner and outer GC subsamples (both with S/N $\sim25$, and 15 and 29 spectra, respectively, as visualized in the middle and bottom plots of Fig.~\ref{fig:GCstacksSN}). 
When stacking all the GCs, inner and outer, the highest S/N of 36.24 is attained by stacking 37/49 GC spectra.
For the inner GCs only, we stacked 15/20 GC spectra, because this results in the highest attainable S/N of 25.36 for the stacked spectrum. For the outer GCs, stacking all of the 29 outer GC spectra leads to the best S/N of 26.26.  

Finally, we also selected and stacked the spectra of the two innermost GCs, only considering distances $<1.5$ kpc. 
In this region, \citet{Barbosa2021} found that the stellar light was dominated by the CP (or relic) formed through a first very short, high-z star-formation episode. 
However, given the much smaller number of objects stacked together, the final spectrum has a limited S/N ($\sim6$). Luckily, some of the main stellar absorption features are still visible, and hence a tentative inference on the age and metal content could be obtained. 

    

%
The entire wavelength range of the final stacked spectra of all (top), inner (middle), and outer (bottom) GCs can be seen in Figure~\ref{fig:GCstacksspectra}. 
The dashed vertical lines show some of the most prominent stellar absorption features in the optical and NIR coming from hydrogen, magnesium, calcium, iron, and sodium. Below the spectra, in each of the panels,  we plot the corresponding error spectra in blue. Noise peaks corresponding to the atmospheric emission line of atomic oxygen near $5580\,\AA$ \citep{Deutsch2003}, and redder telluric H$_2$O lines are clearly visible from it.

\section{Stellar population analysis}
\label{sec:stelpop}

The aim of this section is to determine the ages, metallicities, and $\alpha$-enhancements of the stacked GC spectra. 
In particular, the goal of producing and characterizing the stellar populations of the inner and outer GCs separately is to assess whether the different color distribution also implies different stellar population properties, such as a different integrated age and or a different metallicity and/or [$\alpha$/Fe] ratio. Despite the lower S/N, we also attempted to measure stellar population parameters for the spectrum produced stacking only the two innermost GCs, trying to detect differences with the other GC populations and hence identifying GCs bound to the core progenitor of NGC~3311. 

To perform the stellar population analysis on the co-added GC spectra, we used a full spectral fitting approach with the python package {\tt pPXF}, which stands for penalized PiXel-Fitting. Originally developed by \citet{Cappellari2004}, and then substantially improved in \cite{Cappellari2017}, this method uses the maximum penalized likelihood to estimate a robust fit to an input spectrum based on a given library of SSP models. The best fit consists of a weighted combination of the input SSPs, which enables one to estimate the properties of the population producing the spectrum.

Since we are also interested in [$\alpha$/Fe], which, as proxy for [Mg/Fe]\footnote{We note, however, that  [$\alpha$/Fe] and [Mg/Fe] are not exactly the same. For instance, in Fig.~9 of \citet{Barbosa2021} one can see that a slightly super-solar [$\alpha$/Fe] can result from a combination of a larger [Mg/Fe] (indicating quick star formation) plus roughly solar or slightly subsolar [Ti/Fe] and [Ca/Fe].}, contains important information about the timescale of the star formation \citep{Thomas2005}, we customized the original version of {\tt pPXF} to let it work in a 3D space: age, metallicity, and [$\alpha$/Fe]. 
As templates for the fit, we used the new sMILES models \citep{Vazdekis23} keeping the IMF fixed (bimodal with $\Gamma=1.3$), as we limited the fitting to wavelengths bluer than 6750\AA, where the dependency on IMF is minimal. We tested the robustness of the age and metallicity results, however, by extending the fit region to also cover the CaT lines. The inferred parameters remain unchanged (within 1$\sigma$ errors). In particular, the SSP model grid that we input into the fitting code is composed of 53 different ages, ranging from 0.03 to 14.0 Gyr, 
ten different metallicities ranging from -1.79 to +0.26, and five different [$\alpha$/Fe] ratios ranging from -0.2 to 0.6 in steps of 0.2. 

We performed a {\tt pPXF} fit  on the stellar component only, since gas should be negligible in GCs. 
We first rebinned the GC spectral data into log-space and set up all necessary parameters, such as the initial estimates for the velocity and velocity dispersion\footnote{Since the spectra have been de-redshifted before stacking, we used an initial velocity estimate of zero.}, the full width at half maximum (FWHM) of the templates and the GC spectra, and the degree of the multiplicative polynomial used in the fit to correct for the shape of the continuum. 
In order to obtain reliable stellar population parameters, it is best to run the fit without using additive polynomials, since they might alter the width of some features and consequently distort the results on the stellar population parameters. We therefore only used a multiplicative polynomial (MDEGREE) of grade=7 to correct for possible imperfections in the continuum shape\footnote{In \citet{Spiniello21b}, a test was performed showing that MDEGREE values between $\sim5$ and $\sim15$ produce stable results on spectra of similar S/N. We have indeed verified that using MDEGREE=14 produce only negligible differences in the estimated stellar population parameters}. We performed the fit without any regularization (see the code documentation for more information) as it is possible to find GCs approximately with a single stellar population.

\begin{figure*}[]
        \centering
        \begin{subfigure}[b]{\textwidth}
            \centering
            \includegraphics[width=0.95\textwidth]{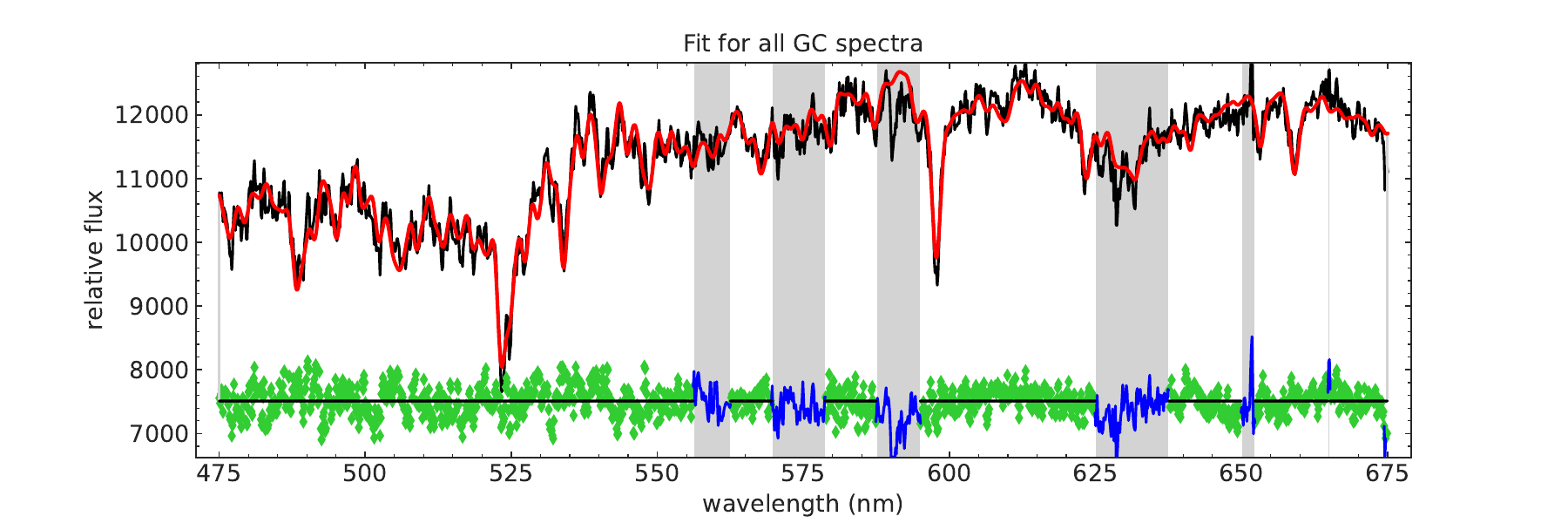}
            \end{subfigure}
        \hfill
        \begin{subfigure}[b]{\textwidth}  
            \centering 
            \includegraphics[width=0.95\textwidth]{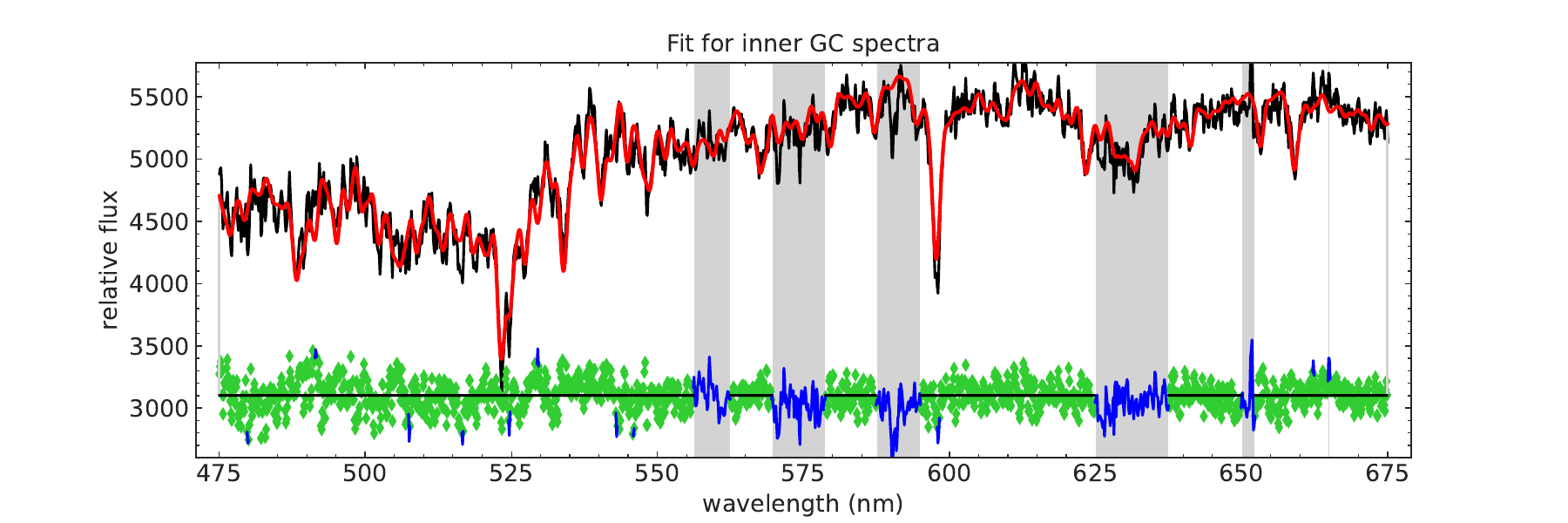} 
        \end{subfigure}
        \hfill
        \begin{subfigure}[b]{\textwidth}   
            \centering 
            \includegraphics[width=0.95\textwidth]{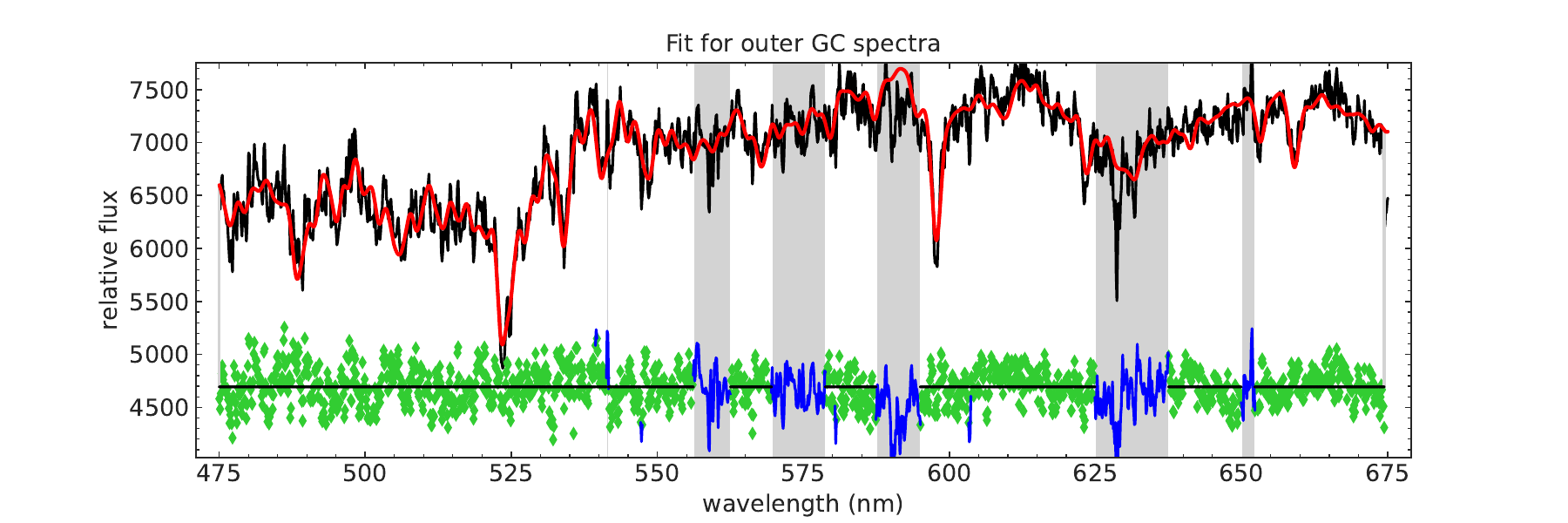}
        \end{subfigure}
        \hfill
        \begin{subfigure}[b]{\textwidth}   
            \centering 
            \includegraphics[width=0.95\textwidth]{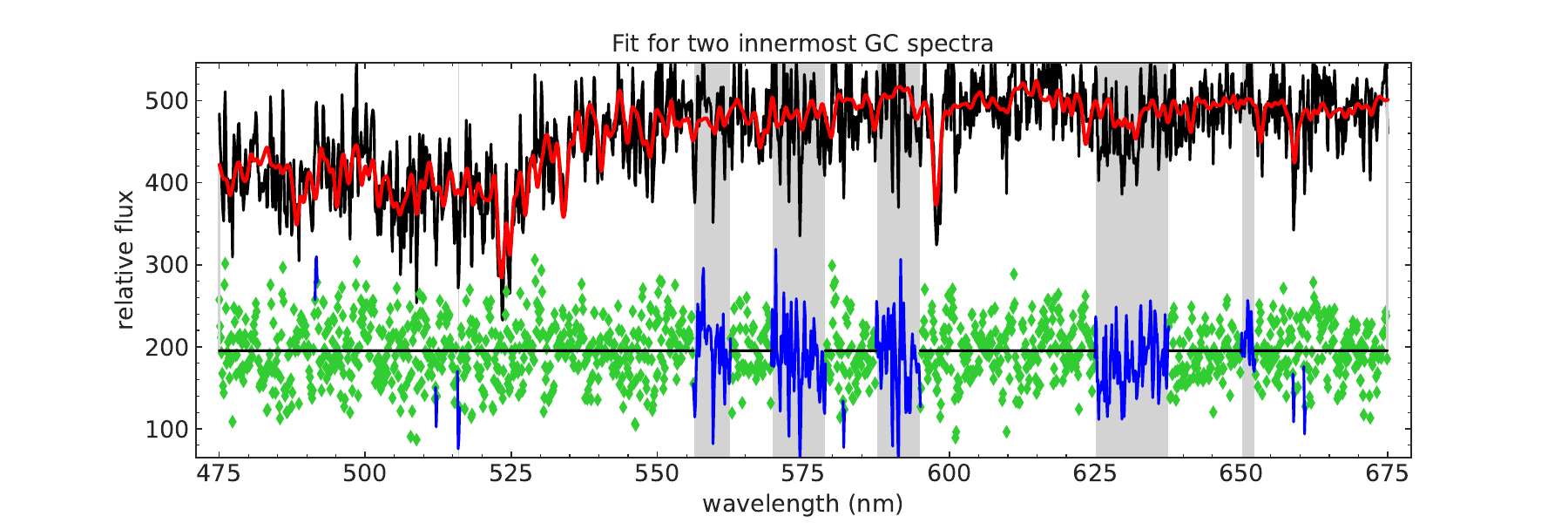}
        \end{subfigure}
        \caption{ \texttt{pPXF} fit for stacked spectra of all (top), inner (second), outer (third), and two inner (bottom) GCs. The input spectra are shown in black, the fits in red, and the residuals in green. The gray vertical bands and blue points in the residuals indicate regions that were excluded from the fitting process.} 
        \label{fig:ppxf}
    \end{figure*}

First, we ran an initial fit on the data, from which we obtained an estimate of the scaling for the noise array. This scaling allowed us to account for the fact that uncertainties might be overestimated or underestimated. The noise scale factor was set to make the residual distribution look like a standard normal distribution. In the second run of the fit, we re-scaled the noise by this factor and also removed significant outliers from the fit with the CLEAN keyword argument as well as manually in order to obtain an improved result. 
The fits for the four stacked spectra (all, inner, outer, and two innermost GCs) are shown in Figure~\ref{fig:ppxf}. We finally inferred mass-weighted ages and metallicities from the weights of the SSP models used by the fit using Equations (1) and (2) from \citet{McDermid2015}. 

Since {\tt pPXF} does not directly determine errors on the age, metallicity, and [$\alpha$/Fe], we applied an empirical and statistical approach in order to obtain estimates on the uncertainties. In this process, we first determined the best-fit model and the residuals with {\tt pPXF}. 
Then we randomly reshuffled the order of these residuals and add them back to the best-fit model to produce a new, slightly perturbed spectrum. We note that this approach assumes a Gaussian distribution of the uncertainties, however, without requiring any prior assumption on them. 
We also considered the fact that some parts of the spectrum are noisier than others. To do this, we first divided the residuals by the noise in order to remove the wavelength dependence from it, and only then did we reshuffle the residuals. Finally, we scaled the reshuffled array back by multiplying it by the noise. In this way, we perturb the fit so that it is consistent with the noise properties of our data. We then ran {\tt pPXF} to determine the age, metallicity, and [$\alpha$/Fe] of the perturbed spectrum. This procedure was repeated 100 times, each time with newly reshuffled residuals. We finally estimated the uncertainties on the properties by taking the standard deviation of the 100 resulting values. 

The results of the analysis are listed in Table~\ref{tab:ppxf_results}. 
All of the stacked GC populations are consistent with an old age of $>13.5$ Gyr, but the inner GCs are more metal rich and have slightly lower $\alpha$ abundances than the outer GCs. 
This result 
confirms the differences in color that can be seen in Figure~\ref{fig:sdss_r-i}. As expected, the stack of all GCs has properties in between the outer and inner ones. 

For the co-added spectrum resulting from the two innermost GC candidates only, the stellar population has super solar metallicities [M/H] $= 0.21\pm0.04$, and it is $\alpha$ enhanced with $[\alpha/\text{Fe}] = 0.26\pm0.06$, indicating a very intense and short star-formation episode that gave rise to the stars in these GCs, which are therefore compatible with being bound with the core progenitor. Somewhat surprisingly, the inferred age is slightly younger than, although compatible with, those estimated from the other spectra. However, these quantities have larger uncertainties, given the lower S/N of the spectroscopic data, and are all consistent within one-sigma errors. 

In Appendix~\ref{app:indices}, we check the results obtained via full spectral fitting against these obtained via line-index analysis. We confirm that all the stacked spectra are consistent with very old ages from the [MgFe]--H$\beta$ plot and that the spectrum for the inner population is more metal rich and $\alpha$ poor than that for the outer from the <Fe>--Mgb one. We remind the reader that with line indices one obtains light-weighted quantities,  while {\tt pPXF} produces mass-weighted ones. As expected, the stack of all the spectra lies in between the inner and outer ones in terms of age and metallicity. Using line indices is a helpful tool to confirm our results, since they are less prone to degeneracy. While improvements are possible in the future, especially by extending spectra to longer wavelengths, we believe the current results make the case for the presence of $> 13$ Gyr old stellar populations in the bona fide GCs selected in NGC~3311.

\begin{table}[t!]
    \centering
    \caption{Age, metallicity, and [$\alpha$/Fe] of the stacked GC spectra.}
    \begin{tabular}{cccc}
    \hline
    \hline
    Population     & Age (Gyr) & [M/H] & [$\alpha$/Fe] \\
         \hline
    all &  $13.89\pm 0.40$ &$ 0.15 \pm 0.02 $ & $0.17 \pm 0.03 $\\
    inner & $13.93 \pm 0.40$ &  $ 0.19 \pm 0.03$ & $ 0.12 \pm 0.03$ \\    
    outer & $13.86\pm 0.60$  & $0.12\pm 0.04$  & $0.18\pm 0.04$\\    
    2 innermost & $13.56 \pm1.24$ & $0.21\pm0.04$ & $0.26\pm0.06$\\
    
        \hline
    \end{tabular}
        \begin{tablenotes}
      \item \textbf{Notes.} These mass-weighted properties were determined for the stacked spectrum for all, only inner, only outer, and only the two innermost GCs of NGC~3311 through an unregularized {\tt pPXF} fit, while uncertainties were estimated through a statistical Monte Carlo approach.
    \end{tablenotes}
    \label{tab:ppxf_results}
\end{table}

\section{Discussion and conclusions}
\label{sec:conclusion}

In this paper, we describe our study of the GC populations in the central region of NGC~3311, the BCG of the Hydra I cluster of galaxies. 
We used MUSE Wild-Field mode Integral field data (IFS), acquired in four pointings: three aligned along the photometric major axis of the galaxy, and one targeting HCC007, a spectroscopically confirmed member of the cluster. From the reduced IFS cubes, we extracted all the point-like sources after modeling and subtracting the light from the central galaxy. Then, we applied a morphological ($\epsilon\leq0.3$), a kinematical (radial velocities between 2500 and 5000 km/s), and a spectroscopical (no emission lines) selection. 
Out of the $680$ point-like sources, $369$ passed the morphological and magnitude criteria, and $124$ had suitable line-of-sight velocity to be bound to NGC~3311. Of these, $74$ display emission lines in their spectra, which characterize them as background objects. Hence, the simultaneous application of all the above criteria led to the identification of $49$ objects, classified as bona fide GC candidates. The vast majority of them are contained in field A, the one centered on NGC~3311. 
    
In order to build spectra with a sufficiently high S/N to constrain the GC stellar populations, we co-added the 1D spectra of single GCs together. We did that for the entire population of reliable candidates (37/49 of them, which led to the highest possible S/N of $36.24$), and for two subsamples, inner GCs (15 spectra, resulting S/N=$25.36$), and outer GCs (29 spectra, resulting S/N=$26.26$), respectively. 
The separation is carried out at a distance of d=20 arcsec, corresponding to $\sim 4$ kpc, from the NGC~3311 center. In addition, we co-added only the spectra of the two innermost GCs with $d<1.5$ kpc, the region where the light is dominated by the core progenitor of NGC~3311 \citep{Barbosa2021}. 

We then obtained the SDSS $r-i$ color distribution of the inner and outer GC populations, finding that, although the colors are similar, there is an overall shift of the inner GCs toward redder colors. Furthermore, we constrained the mass-weighted age, metallicity, and [$\alpha$/Fe] of all the GCs; we also did this for the inner, outer, and two innermost GCs. We used the code {\tt pPXF}, which performs a full-spectral-fitting analysis comparing the spectra with a large grid of SSP models with varying age, metallicity, and [$\alpha$/Fe]. We kept the IMF fixed, but we restricted the fit to wavelengths bluer than 6750\AA in order to minimize the IMF dependency in the spectral features we considered. 

The main results of the stellar population analysis on the co-added spectra of bona fide GCs are as follows. 
\begin{itemize}
    \item[i)] Both the inner and outer GC populations, as well as the two innermost ones, have very old and consistent ages ($>13.5$ Gyr). 
    \item[ii)] The inner GCs are slightly more metal rich ([M/H] $=0.19\pm0.03$) than the outer ones ([M/H]$=0.12\pm0.04$), as already hinted by the redder $r-i$ colors of the inner GC subsample. 
    \item[iii)] All spectra are consistent with having an $\alpha$-enriched population, but the outer GCs have a higher [$\alpha$/Fe] ratio than the inner ones ($0.18$ versus $0.12$ for the inner). 
    \item[iv)] The two innermost GCs show slightly younger ($13.6\pm1.2$) but consistent ages. They have the highest metallicity  ([M/H]$=0.21\pm0.04$) and [$\alpha$/Fe] ($0.26\pm0.06$) population. However, since the stacked spectrum of these two GCs has a lower S/N, these results, especially for the age, have larger uncertainties. 
    \item[v)] The parameters inferred for the stacked spectrum for the entire population lie in between the inner and outer subsamples, as expected.
\end{itemize}

These stellar population results are fully consistent with the spatially resolved ones obtained in \citet{Barbosa2021}, where they found that the innermost light is dominated by stars belonging to the preserved CP. In fact, from the bottom panel of their Fig.~7, one can clearly see that the stellar population parameters are rather flat within the inner 3 kpc, with very old ages (12 Gyr), super-solar metallicities ([M/H]$\sim0.2$), and only slightly enhanced in $\alpha$ elements ([$\alpha$/Fe]$\sim0.1$). The results for the two innermost GCs are compatible with a very intense star formation on a very short timescale, which supports the claim that the CP in NGC~3311 was not destroyed after its in situ formation.

At larger distances from the center of NGC~3311, the stellar age and metallicity decrease, while the [$\alpha$/Fe] and the stellar velocity dispersion rise. Although we did not cover a very large distance with the GCs, we observe a similar behavior: the stars in the outer stack have 
lower metallicities and higher [$\alpha$/Fe]. However, the age is still old and consistent with that inferred from the inner stack. 
Hence, at distances $\ge 4$ kpc, the GCs start to be associated with the photometric component "C" identified in \citet{Barbosa2018}, which, differently from the entire galaxy classified as slow rotator, shows signs of a regular rotation pattern. Unfortunately, we do not identify any GCs associated with the cD envelope and the large halo, where the stellar populations are much younger (7-8 Gyr) and relatively more metal poor ([M/H]$\sim 0$).

Finally, our results on the stellar population ages of the bona fide GC do not support the previous claim by  \citet{Hempel05} of the presence of a substantial fraction ($50$ to $60\% $ in number of their selected sample of unresolved sources) of intermediate (to young) populations of GCs. We do find that a large fraction of the unresolved sources around the center of NGC~3311 in the selected line-of-sight velocity range have detectable emission lines, and we classified these as background objects in this work. They may be star forming satellites, perhaps irregular or dwarf galaxies, which are infalling toward the BCG of the Hydra I cluster, and they will be part of a subsequent study.

Although GCs do not necessarily follow the distribution of the underlying stellar populations, our main results would be unchanged if using a different separation between the inner versus outer sample, because there appears to be no intermediate-age population. The chosen separation was motivated by the photometric components found in \cite{Barbosa2018} (see their Table 1) and the fact that this led us to approximately similar subsamples for the inner and outer GCs. Finally, GCs are a population embedded in the gravitational potential of NGC~3311. By adopting a 20 arcsec radius, we are effectively comparing those within and outside the half-light radius of the central galaxy.

The current analysis demonstrates the importance of obtaining high-quality spectroscopic data, such as those produced by the MUSE instrument, for the identification and the study of the age and metallicity of GCs. High S/N spectroscopy is the robust way to separate bona fide GCs from compact/background objects with emission lines.
In the near future, thanks to a new, ongoing MUSE Large Program (LEWIS, see \citealt{Iodice2023}), more spectra of GCs associated with the Hydra I cluster will become available at much larger galactocentric distances from NGC~3311. It will therefore be possible to extend this work and study the population of GCs associated with the BCG, ultra-diffuse galaxies in the cluster (the main targets of LEWIS), and, even more interestingly, the population of intra-cluster and halo GCs. 

\begin{acknowledgements}
N. Grasser, M. Arnaboldi and L. Coccato would like to acknowledge support from the ESO Summer Studentship program. This research has made use of data from the Multi Unit Spectroscopic Explorer \citep{MUSE2010} from the European Southern Observatory (ESO). Furthermore, this research was made possible thanks to the use of Python (\url{https://www.python.org/}), in particular Astropy \citep{Astropy}, a community-developed core Python package for astronomy, NumPy \citep{Numpy} and Matplotlib \citep{Matplotlib}.
\end{acknowledgements}

\bibliographystyle{aa}
\bibliography{gc.bib}

\appendix
\section{Line-index analysis}
\label{app:indices}
In this appendix, we show the results obtained from line-index analysis on the stacked spectra for the inner, outer, and all GCs. In particular, we measured equivalent widths of the H$\beta$, the Mg$b$ doublet at 5177\AA, and nine different iron lines (Fe4920, Fe5015, Fe5270, Fe5335, Fe5406, Fe5709, Fe5782, Fe6189, Fe6497) in the wavelength region covered by the MUSE data. 
We calculated the line strengths of these lines with the code SPINDEX from \citet{Trager08} for both the GCs spectra and SSP models brought to the same resolution of $300$ km/s. Consistently with the main body of the paper, we used the sMILES models \citep{Vazdekis23} with a bimodal IMF with $\Gamma=1.3$, age of 6, 8, 10, and 12 Gyr, metallicities ranging from -0.25 to +0.4, and [$\alpha$/Fe] ranging from 0 to 0.4. We also linearly interpolated the publicly available models to build a finer grid with a step of 0.1 in [$\alpha$/Fe].

\begin{figure}
    \centering
\includegraphics[width=\columnwidth]{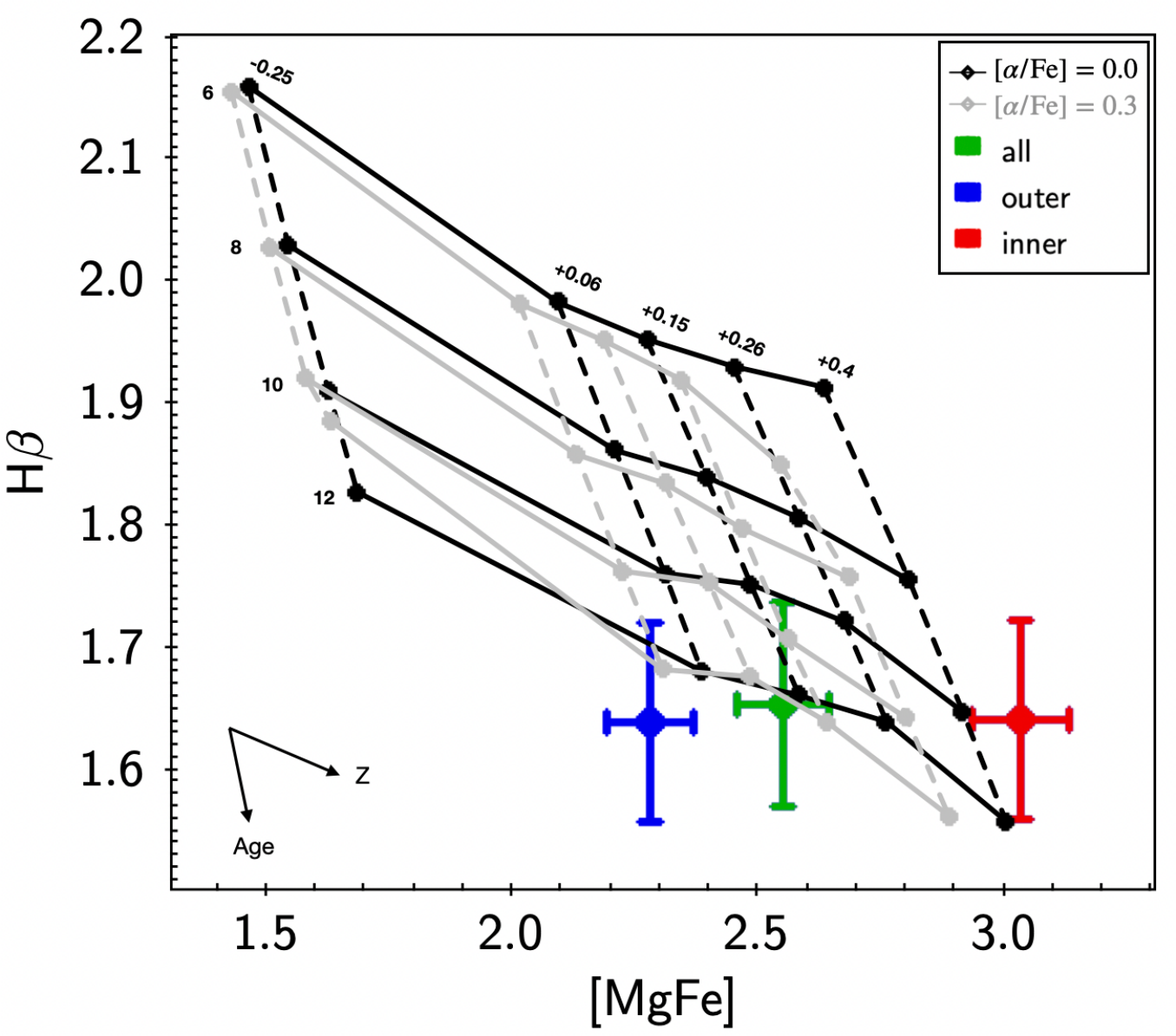}
    \caption{[MgFe] -- H$\beta$ index--index plot used to infer light-weighted age and metallicity. The direction of variation for these two parameters is given by the arrows in the bottom left corner.}
    \label{fig:index-index1}
\end{figure}

\begin{figure}
    \centering
\includegraphics[width=\columnwidth]{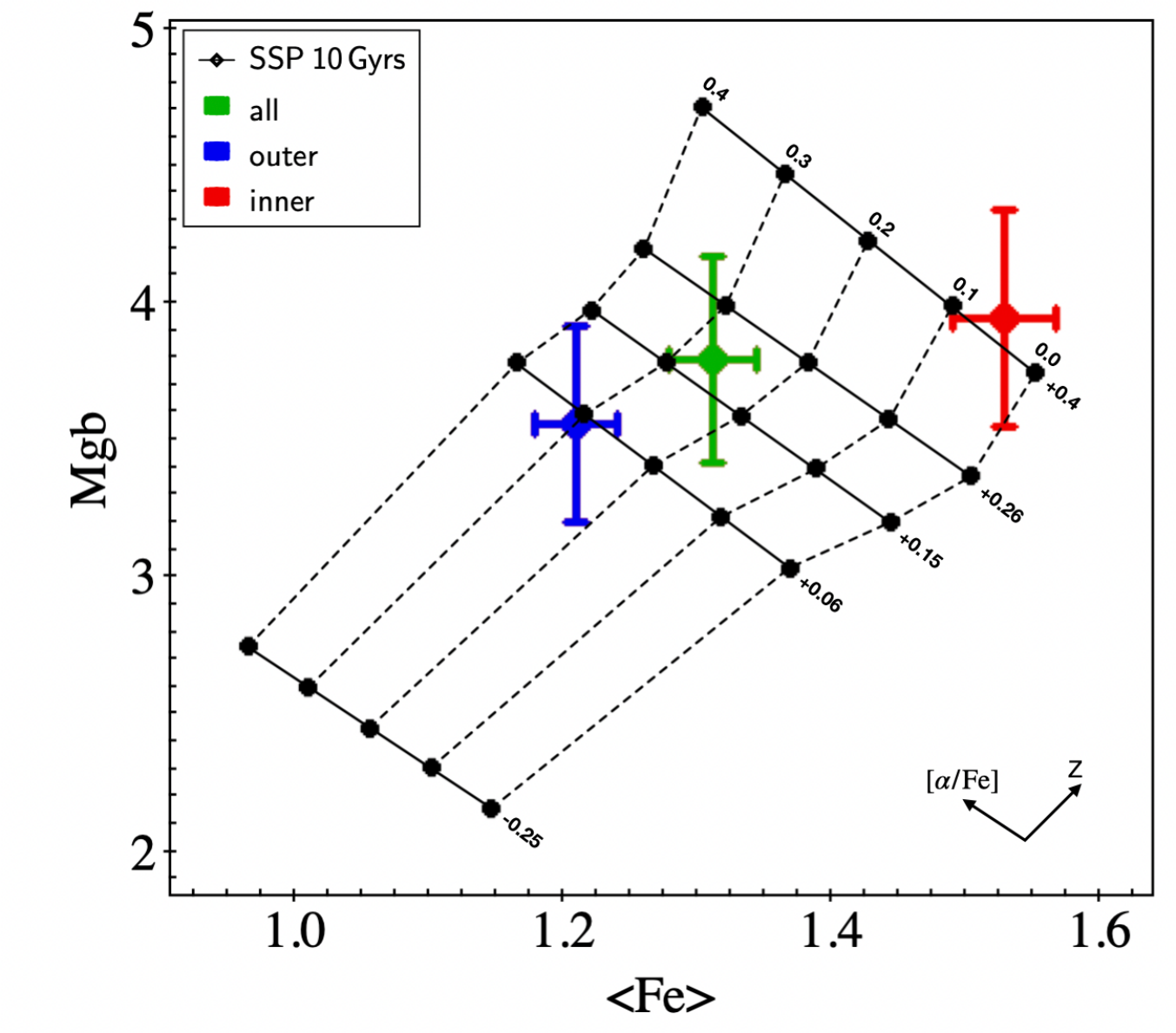}
    \caption{Mg$b$ -- <Fe> index--index plot used to infer light-weighted [Mg/Fe] on GC stacked spectra. The grid shows MILES SSPs with  age of 10 Gyr, covering a range of metallicities  and [$\alpha$/Fe] values. The direction of variation for these two parameters is given by the arrows in the bottom-right corner. }
    \label{fig:index-index2}
\end{figure}

The result of the line-index analysis can be seen in  Figure~\ref{fig:index-index1} and Figure~\ref{fig:index-index2}. 
The first figure shows the [MgFe]\footnote{The index is defined as [MgFe]=$\sqrt{\mathrm{Mg}b}\,\times$ <Fe>, where the <Fe> is the average between the 9 Fe lines. } versus H$\beta$, which is often used to constrain metallicity. In this plot, the [$\alpha$/Fe] only play a marginal role, while age and metallicity vary in orthogonal directions, as indicated by the arrows in the bottom left corner of the figure. 
The results for outer, inner, and all GCs are shown as colored points. They are all consistent with a very old stellar population ($>10$ Gyr), which is perfectly consistent with the results inferred from full-spectral fitting. The inner stack is clearly metal richer than the outer one, also confirming the results presented in the main body of the paper. We note that the quantities inferred by line-index analysis are light-weighted, while the ones constrained via full-spectral fitting are mass-weighted; hence, it is not surprising that the inferred quantities do not fully agree in an absolute sense. 

Figure~\ref{fig:index-index2} shows the Mg$b$ -- <Fe> index--index plots, where the [$\alpha$/Fe] varies in an orthogonal direction to metallicity (see the arrows in the bottom right corner), while age plays only a very marginal role. Here, we only show models with an integrated age of 12 Gyr, given the results we obtained. 
The plot indicates that the inner GCs have a lower [$\alpha$/Fe] ($\sim 0$-$0.1$) with respect to the value inferred for the outer GCs ($\sim 0.3$), once again in good agreement with what was obtained from a full spectral fitting.

\end{document}